\documentclass[aps,reprint,groupedaddress,amssymb,nofootinbib,prb,nobibnotes]{revtex4-1}
\usepackage[hidelinks]{hyperref}
\usepackage{xcolor}
\usepackage{colortbl}
\hypersetup{
	colorlinks,
	linkcolor={red!50!black},
	citecolor={blue!50!black},
	urlcolor={blue!80!black}
}
\usepackage{graphicx}
\usepackage{graphics}
\usepackage{amsmath} 
\usepackage{subcaption}

\usepackage{dcolumn}
\usepackage{bm}
\usepackage{multirow}
\usepackage{soul}
\usepackage{wasysym}

\begin{document}
	
	
	\title{Proton-transfer dynamics in ionized water chains using real-time Time Dependent Density Functional Theory}
	
	
	\author{Vidushi Sharma}
	\email{vidushi.sharma@stonybrook.edu}
	\author{Marivi Fern\'{a}ndez-Serra}
	\email{maria.fernandez-serra@stonybrook.edu}
	\affiliation{Department of Physics and Astronomy, Stony Brook University, Stony Brook, New York 11794-3800, United States}
	\affiliation{Institute for Advanced Computational Science, Stony Brook University, Stony Brook, New York 11794-5250, United States}
	
	
	\date{\today}
	
	\begin{abstract}
		In density functional-theoretic studies of photoionized water-based systems, the role of charge localization in proton-transfer dynamics is not well understood.
		This is due to the inherent complexity in extracting the contributions of coupled electron-nuclear non-adiabatic dynamics in the presence of exchange and correlation functional errors.
		In this work, we address this problem by simulating a model system of ionized linear H-bonded water clusters using real-time Time Dependent Density Functional Theory (rt-TDDFT)-based Ehrenfest dynamics.
		Our aim is to understand how self-interaction error in semilocal exchange and correlation functionals affects the probability of proton transfer.
		In particular, we show that for H-bonded (H$_2$O)$_n^+$ chains (with $n>3$), the proton-transfer probability attains a maximum, becoming comparable to that predicted by hybrid functionals.
		This is because the formation of hemibonded-type geometries is largely suppressed in extended H-bonded structures.
		We also show how the degree of localization of the initial photo-hole is connected to the probability of a proton-transfer reaction, as well as to the separation between electronic and nuclear charge.
		These results are compared to those obtained with adiabatic dynamics where the initial wavefunction is allowed to relax to the ground state of the ion cluster, explaining why different functionals and dynamical approaches lead to quantitatively different results.
	\end{abstract}
	
	
	\maketitle
	
	
	\section{Introduction}
	Understanding charge transfer dynamics in aqueous solvated semiconductor surfaces for enhanced photocatalytic activity is a complex undertaking. 
	Elucidating the structure of these hydrated surfaces has been a subject of extensive research not only because of the ubiquitous nature of water but also to better control and improve the (photo)catalytic property of the adherent semiconductor materials.
	Experimentally, even the characterization of the structural motifs and physisorbed chemical species on the semiconductor surfaces is both difficult and uncertain; for example, the identification of surface species involved in photocatalysis is often based on indirect evidence \cite{C8CP07632D}.
	In particular, the distinction among oxygen atoms belonging to a (surface-bound) hydroxyl group, an oxidated surface and a peroxide compound is, at best, qualitative.
	Such species are intermediates in the half water-splitting oxidation reaction \cite{doi:10.1021/jp102958s}, and their identification is linked to a mechanistic understanding of how photo-excited holes are transferred to the surface-adsorbed species and how protons move away from the reaction sites.
	In general for water-oxide interfaces, the identification of reactive species -- electrons and holes -- and their distribution in the bulk versus that at the surface, pose an uphill challenge for uncovering the underlying reaction mechanisms \cite{doi:10.1021/jp031305d, doi:10.1021/ja8034637}.
	
	From a theoretical viewpoint, \textit{ab initio} molecular dynamics studies have led to a significant understanding of dynamical processes at the (photo)catalytic interface \cite{Brini2017, C9SC05116C, C9CP05097C, doi:10.1021/jacs.6b03446, C6SC04378J}.
	These include a description of the state of adsorbed water -- molecular or dissociated -- favored on a TiO$_2$ anatase surface \cite{PhysRevLett.81.2954, doi:10.1021/jp037685k}, as well as the facet-dependent behavior of excess electrons that makes one surface-termination better suited for a reduction reaction and the other better suited for oxidation \cite{Selcuk2016}.
	There remain even more unresolved puzzles. For instance, whether or not the transfer of a (photogenerated) hole to surface-water species leads to a fast separation of the proton remains an open question.
	Furthermore, the time scale of the proton-transfer reaction and whether it ought to be studied at the adiabatic or non-adiabatic level of theory \cite{Yuan19148, doi:10.1021/ar100153g, C3CP51440D, doi:10.1063/1.3664746, Richter2018} is also still an open question. 
	These questions are difficult to attack largely due to the inherent complexity of the systems involved, in particular due to large system sizes, the necessity for long simulation times, and potential contributions from non-adiabatic terms.
	
	A principal aim of our work is to develop physical intuition for a more tractable system comprising of water molecules and examine proton-transfer probability (P(PT)) as computed by rt-TDDFT-based \cite{Ullrichbook, Maitra_2017} Ehrenfest dynamics \cite{Ehrenfest1927, doi:10.1063/1.471952, Miyamoto2015, PhysRevA.56.162} upon ``photoionization'' of such systems. The main idea behind this study is to identify a practical and reliable method to study more complex systems.
	In the condensed phase, water molecules are organized in large cluster structures, strongly connected by hydrogen-bond networks \cite{Stillinger451, PhysRevLett.96.016404}. Frequently (aqueous) solvated semiconductor surfaces sustain and enhance the formation of such networks \cite{doi:10.1021/jp302793s, doi:10.1021/jp037685k, C8SC03033B, doi:10.1021/ja403850s, doi:10.1021/acs.jpclett.7b00358}, resulting in the ordered formation of surface chains of water molecules. 
	Thus the very well-developed network of H-bonds in water plays a fundamental role in not only defining its structural and dynamical properties, but also the properties of the material it interacts with \cite{PhysRevLett.81.1271, D0SC01517B}.
	
	We also want to understand the effect of hydrogen bonds and their cooperativity -- if any -- on hole localization and proton transfer rate, and simultaneously evaluate
	the reliability of a commonly used semilocal generalized-gradient approximate (GGA) exchange and correlation (XC) density
	functional, namely PBE \cite{PhysRevLett.77.3865}, on describing these processes. 
	We focus on open-ended chains of water molecules such that upon photoionization, the hole is always localized at the first H-bond donor 
	oxygen of the chain. 
	Since removing an electron from a system comprising of $n$ H$_2$O molecules makes it extremely prone to self-interaction error (SIE) by conventional standards \cite{doi:10.1021/acs.jpclett.8b00242}, we first address this issue before conducting a (TD)DFT-based dynamics of ionized water clusters. 
	We compute the ground-state static binding energies of H$_2$O in ionized water clusters, (H$_2$O)$_n^+$, using various approximate (semilocal and hybrid) density-functionals and higher accuracy wavefunction-based methods.
	A comparison of these independent binding energies yields an estimate of the underlying SIE inherent to the XC functional used.
	SIE manifests in the deviation of the binding energy computed using a given XC functional from its value using a more accurate method. 
	We show that SIE in functionals like PBE (GGA) become less severe for condensed phases.
	This is a crucial observation because most assessments of SIE in PBE among other XC functionals have targeted smaller molecular systems or systems with non-interacting components \cite{10.1093/nsr/nwx111, doi:10.1063/1.2566637, doi:10.1021/jp011239k}. 
	The first part of our work, outlined in Sec. \ref{sec:sie_waterion}, addresses this gap by examining the error for an interacting system, (H$_2$O)$_n^+$, which is closer in nature to the bulk phase.

	In Sec. \ref{sec:molecular_dynamics}, we present simulations of the dynamics of (H$_2$O)$_n^+$ systems using (TD)DFT techniques. 
	Our aim is to understand the proton-transfer mechanism and identify the relevant time scales. 
	We present an rt-TDDFT/PBE-based Ehrenfest trajectory statistics of the proton-transfer in H-bonded ionized-water chains of size $n = (2-5)$.
	We also compare the non-adiabatic Ehrenfest dynamics with adiabatic Born-Oppenheimer molecular dynamics, and further refine the analysis by distinguishing between PBE and the hybrid functional PBE0 \cite{doi:10.1063/1.472933, doi:10.1063/1.478522}.

	Finally in Sec. \ref{sec:hole_dynamics}, we explore the dynamical evolution of the photo-hole in the ionized water cluster.
	The rt-TDDFT/Ehrenfest approach involves a simultaneous real-time evolution of the electronic subsystem in accordance with the time-dependent Kohn-Sham equations and a classical evolution of nuclei.
	We investigate explicitly the evolution of hole densities and establish a connection between hole-localization and initiation of proton-transfer dynamics in ionized water chains.
	We also compare the behavior of three density functionals (with increasing fractions of exact exchange) for calculations of the dynamic hole densities on a particular H$_2$O unit of the ionized water chain.
	Our results indicate that for the larger clusters, PBE simulations are not substantially different from those obtained with hybrid functionals. 
	We explain why PBE produces significantly different results in smaller clusters, with our findings supporting this observation as well as the use of other semilocal XC functionals in rt-TDDFT/Ehrenfest simulations of larger, condensed phase systems.
	\section{\label{sec:sie_waterion}Self-interaction error in ionized water clusters}
	
	\begin{figure*}
		\includegraphics[width=5.8in]{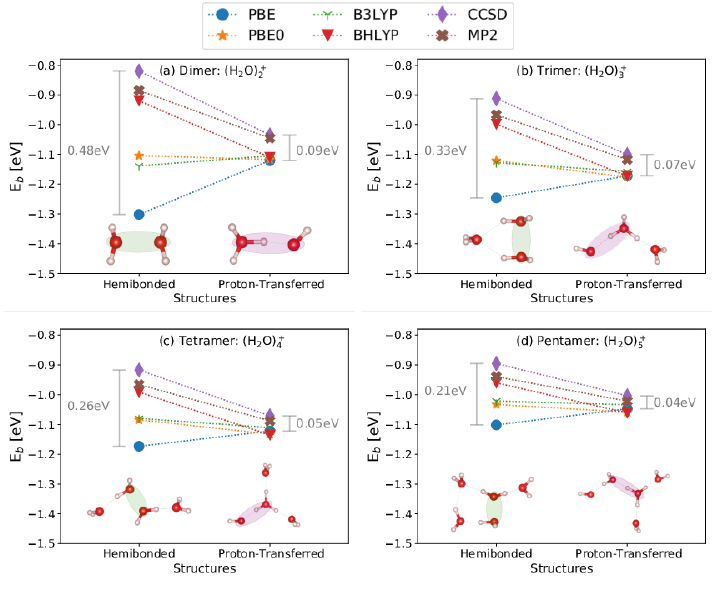}
		\caption{\label{fig:sie}Binding energies of the hemibonded and proton-transferred  geometries of the ionized water clusters (H$_2$O)$_n^+$, for (a) $n{=}2$, (b) $n{=}3$, (c) $n{=}4$, (d) $n{=}5$. The associated atomic structures are shown in the insets. The green (purple) -- shaded regions represent hemibonded (proton-transferred) -- type bonds.}
	\end{figure*}
	Local and semilocal generalized-gradient XC functionals (LDA; GGA) suffer from what is known as self-interaction error (SIE) \cite{Sharkas11283}. 
	The approximation to the exact exchange in these functionals prevents the exact cancellation between the self-Coulomb and self-exchange for all one-electron densities. In order to minimize the total energy, single electron (Kohn-Sham) orbitals tend to over-delocalize their associated electron density. 
	Therefore, for a charged system, the SIE spreads the electron (or hole) artificially over the fragments, yielding too low energies for the delocalized states. 
	The terms ``delocalization error'' and ``self-interaction error'' are often used interchangeably \cite{PhysRevLett.100.146401}, the former typically signifying the physical aspect of the error \cite{doi:10.1021/cr200107z}.

	In this section, we explore the SIE in different charged water clusters, (H$_2$O)$_n^+, (n = 2-5)$. 
	Previous investigations of SIE in DFT have focused on how the charge-delocalization (and hence, SIE) is affected by the size of the system \cite{doi:10.1063/1.4937417, doi:10.1063/1.1926277}. 
	However, most studies have considered non-interacting molecular systems where the size is tuned by simply repeating the non-interacting units separated by some finite distance.
	In such cases, the delocalization error worsens with increase in system-size, as ionization would result in the removal of a fraction of electron charge from all the molecules in the system. 

	In our study of ionized (H$_2$O)$_n^+$ clusters, we also consider the effect of electrostatic interactions among the H$_2$O units on the size-dependent charge delocalization or SIE.
	Here $n$ controls the system size, and Fig \ref{fig:sie} illustrates the SIE for two prominent spatial configurations of (H$_2$O)$_n^+$ - hemibonded (HB) and proton-transferred (PT) for each ($n = 2-5$). 
	The hole (i.e. highest occupied molecular orbital (HOMO) of the un-ionized system) has a very different electronic configuration in these two structures. 
	In the hemibonded structure the hole arises from the antibonding combination of the $1b_1$ orbitals (i.e. oxygen $p_z$ orbitals) of each of the two water molecules.
	By contrast, in the proton-transferred structure the hole is largely localized on the $1b_1$ orbital of the H-donating molecule. 
	This gives rise to varying amounts of charge delocalization and SIE for the same system.
	We compare different methods used for predicting the energies of the hemibonded and proton-transferred (H$_2$O)$_n^+$ clusters.
	These include the semilocal PBE functional, hybrid functionals which combine different amounts of exact Hartree-Fock (HF) and DFT exchange - B3LYP ($E^{HF}_X = 20\%$), PBE0 ($E^{HF}_X = 25\%$), BHLYP ($E^{HF}_X = 50\%$), and post Hartree-Fock methods - M{\o}ller-Plesset perturbation with second-order correction (MP2), and coupled-clusters for singles and doubles (CCSD). 

	For the water dimer cation, (H$_2$O)$_2^+$, Sodupe \textit{et al.}  \cite{doi:10.1021/jp983195u} compared different DFT and post-Hartree-Fock methods to determine its ground state structure. 
	Their observation is that GGA functionals overestimate the energies of the delocalized hole, thus incorrectly favoring the hemibonded configuration as the preferred ground state of the dimer ion. 
	However a hybrid functional combining equal fractions of GGA and exact (Hartree-Fock) exchange -- BHLYP -- improves the hole-localization, and correctly predicts the proton-transferred geometry as the ground-state, in accordance with more accurate wavefunction-based approaches. 
	In the first row of Table \ref{tab:hbptbenergies}, we compare the energies obtained at different levels of theory for the hemibonded and proton-transferred dimer ions; our results are in agreement with previous studies \cite{doi:10.1021/jp983195u, C3CP51440D, doi:10.1021/jp802140c, doi:10.1021/acs.jpca.6b09905}, but these studies have solely focused on the water dimer cation. 

	The ground-state DFT calculations for all the systems are performed with the 6-311++G** basis set.
	The binding energy per water monomer ($E_b$) is evaluated as
	\begin{equation}
	E_b = \frac{E_{(H_2O)_n^+}-(n-1)E_{H_2O}-E_{H_2O^+}}{n} ~,
	\label{eq:one}
	\end{equation}
	for all $n=(2-5)$ ionized water clusters. In Eq. \eqref{eq:one}, $E_{(H_2O)_n^+}$ is the total energy of an ionized ($+1$) cluster with $n$ H$_2$O molecules, $E_{H_2O}$ is the total energy of a neutral H$_2$O monomer and  $E_{H_2O^+}$ is the total energy of a charged ($+1$) H$_2$O monomer.

	A crucial feature of PBE when applied to ionized water clusters is that it favors the H$_2$O$\cdots$OH$_2$ bonding interaction, yielding rather low energies for structures containing H$_2$O monomers bonded via the O's.
	We refer to these as hemibonded-type geometries.
	The binding energies are shown for the simplest case of a water dimer ion (H$_2$O)$_2^+$ in panel (a) of Fig. \ref{fig:sie}. 
	For the hemibonded structure, the XC functionals tested in this work provide very different binding energies ranging from  -1.3 eV for PBE to -0.92 eV for BHLYP.
	PBE distributes the hole density evenly over the two water units over-stabilizing the hemibond configuration.
	As shown, the energies become larger (less negative) upon increasing the fraction of exact exchange in the XC functional. 
	The exact exchange in the hybrid functionals compensates for the artificial hole delocalization introduced by the approximated exchange, bringing the energies closer to the more accurate MP2 and CCSD values.
	While the binding energies computed by CCSD and PBE differ by 0.48 eV (see Fig. \ref{fig:sie}(a)), BHLYP performs much better with a binding energy difference of 0.09 eV compared to CCSD.
	
	In contrast, for the proton-transferred [H$_3$O$^+-^{\bm{\cdot}}$OH] structure, all the density functionals provide very similar binding energies, which are fairly consistent with the energies from higher accuracy theories. 
	The energy computed by PBE is 0.09 eV lower than that given by CCSD. 
	Nonetheless, PBE wrongly selects the hemibonded (HB) structure as the ground-state representation of the ionized water dimer due to the underlying SIE.
	In Table \ref{tab:hbptbenergies}, we compare the difference in binding energies of the HB and PT structures to this effect for different methods. 
	A negative energy difference for any given method implies that the HB structure is the preferred ground-state. 
	It should be noted that PBE yields negative energy differences for all $n$ (as expected), thus favoring a hemibond over a proton-transfer. 
	However, $( E_{b,\text{HB}}-E_{b,\text{PT}})_{\text{PBE}}$ approaches zero from the left as $n$ increases, implying a reduction in the SIE.
	In other words, for larger ionized water clusters PBE provides better estimates of binding energies of HB and PT structures, but it still misidentifies the HB geometry as its ground-state configuration. 
	In the following sections, we will show that this is the main reason why there is a large disagreement between PBE and hybrid XC functionals in the description of the dynamical evolution of the ionized dimer.
	This error decreases with increasing chain length ($n$), as the occurrence of the hemibonded structure is largely suppressed as soon as extended H-bond chains form.
	Both CCSD and MP2 predict the proton-transferred geometries as the ground-state of the respective (H$_2$O)$_n^+$.
	We remark that the XC functional that closely mimics the trend of wavefunction-based methods is BHLYP.
	
	The binding energy differences between PBE and CCSD methods ($\Delta E_{\text{CCSD}\rightarrow\text{PBE}}$) are highlighted for (a) Dimer, (b) Trimer, (c) Tetramer, and (d) Pentamer ionized water clusters in Fig. \ref{fig:sie} for the two structural configurations -- hemibonded and proton-transferred. The respective molecular structures for all the ionized water clusters \cite{doi:10.1021/jp405052g} are also shown in the insets in Fig. \ref{fig:sie}.
	It is observed that $\Delta E_{\text{CCSD}\rightarrow \text{PBE}}$ is greater for the hemibonded geometries as compared to the proton-transferred ones for a given system size ($n$).
	$\Delta E_{\text{CCSD}\rightarrow\text{PBE}}$ has a smaller spread for the proton-transfer structures.
	More importantly, $\Delta E_{\text{CCSD}\rightarrow \text{PBE}}$, which is indicative of the size-dependent SIE, decreases significantly with an increasing (H$_2$O)$_n^+$ system size. 
	Specifically, it varies from 0.48 eV for $n=2$ to 0.21 eV for $n=5$ in the hemibonded structures, and from 0.09 eV for $n=2$ to 0.04 eV for $n=5$ in the proton-transferred structures.
	
	This apparent size-dependence of $\Delta E_{\text{CCSD}\rightarrow \text{PBE}}$ (or SIE), though similar in behavior, has different origins for the two fundamentally different molecular arrangements of the ionized water structures.
	In hemibonded structures, an increase in $n$ results in a systematic smearing-out of the hole over more H$_2$O units (with PBE), thereby lowering the ``localized" hole density on individual fragments and reducing their contributions to the overall self-interaction present in the system.
	On the other hand, increasing $n$ in proton-transferred structures counteracts the ``delocalizing bias" \cite{PhysRevLett.100.146401} of PBE by effectively localizing the added hole in the ionized system. 
	The mostly linear and open-network arrangement of H$_2$O molecules in PT clusters enables the hole to be selectively localized over specific H$_2$O units. 
	At the microscopic level, this is driven by the cooperative behavior of H-bonds in open-chain water geometries.
	The unidirectional H-bonds in these finite $n$-chains strengthen each other such that the HOMO of the neutral system -- (H$_2$O)$_n$ -- is localized on the oxygen of the H$_2$O which exclusively donates an H-bond (and does not accept one). 
	With increasing $n$ the H-bond cooperativity \cite{Guevara-Vela2016, Stokely1301} becomes relevant. 
	This manifests in a reduction of the SIE.
	This crucial feature guides much of our work on the excited-state dynamics of ionized water clusters that is described below.
	
	In the following sections, we discuss both adiabatic dynamics (using Born-Oppenheimer approximation) and non-adiabatic dynamics (using mean-field Ehrenfest approach) of the ionized water chains, that is, (H$_2$O)$_n^+$ (for $n=2-5$).
	Our aim is to capture the ultrafast processes that drive proton-transfer phenomena in ionized water clusters comprising of mostly linear H-bonds. 
	\begin{table}
		\caption{\label{tab:hbptbenergies}Binding energy differences between the hemibonded (HB) and proton-transferred (PT) HOMO-ionized (H$_2$O)$_n^+$ structures computed at different levels of theory. All energy values are reported per water monomer for a cluster containing $n$ H$_2$O molecules, $n = (2-5)$.}
		\begin{ruledtabular}
			\begin{tabular}{ccccccc}
				\multicolumn{1}{c}{$n$}&\multicolumn{6}{c}{$E_{b,\text{HB}}-E_{b,\text{PT}}$ [eV]}\\
				$ $&PBE&PBE0&B3LYP&BHLYP&MP2&CCSD\\ \hline
				2&$-0.18$&$0.01$&$-0.03$&$0.19$&$0.16$&$0.21$\\
				3&$-0.07$&$0.06$&$0.03$&$0.17$&$0.15$&$0.19$\\
				4&$-0.05$&$0.05$&$0.03$&$0.14$&$0.12$&$0.15$\\
				5&$-0.05$&$0.02$&$0.01$&$0.10$&$0.08$&$0.11$\\
			\end{tabular}
		\end{ruledtabular}
	\end{table}

	\section{Methodology}
	
	\subsection{Adiabatic Born-Oppenheimer molecular dynamics}
	\begin{figure*}
		\begin{subfigure}[b]{0.3\textwidth}
			\includegraphics[width=1.3in]{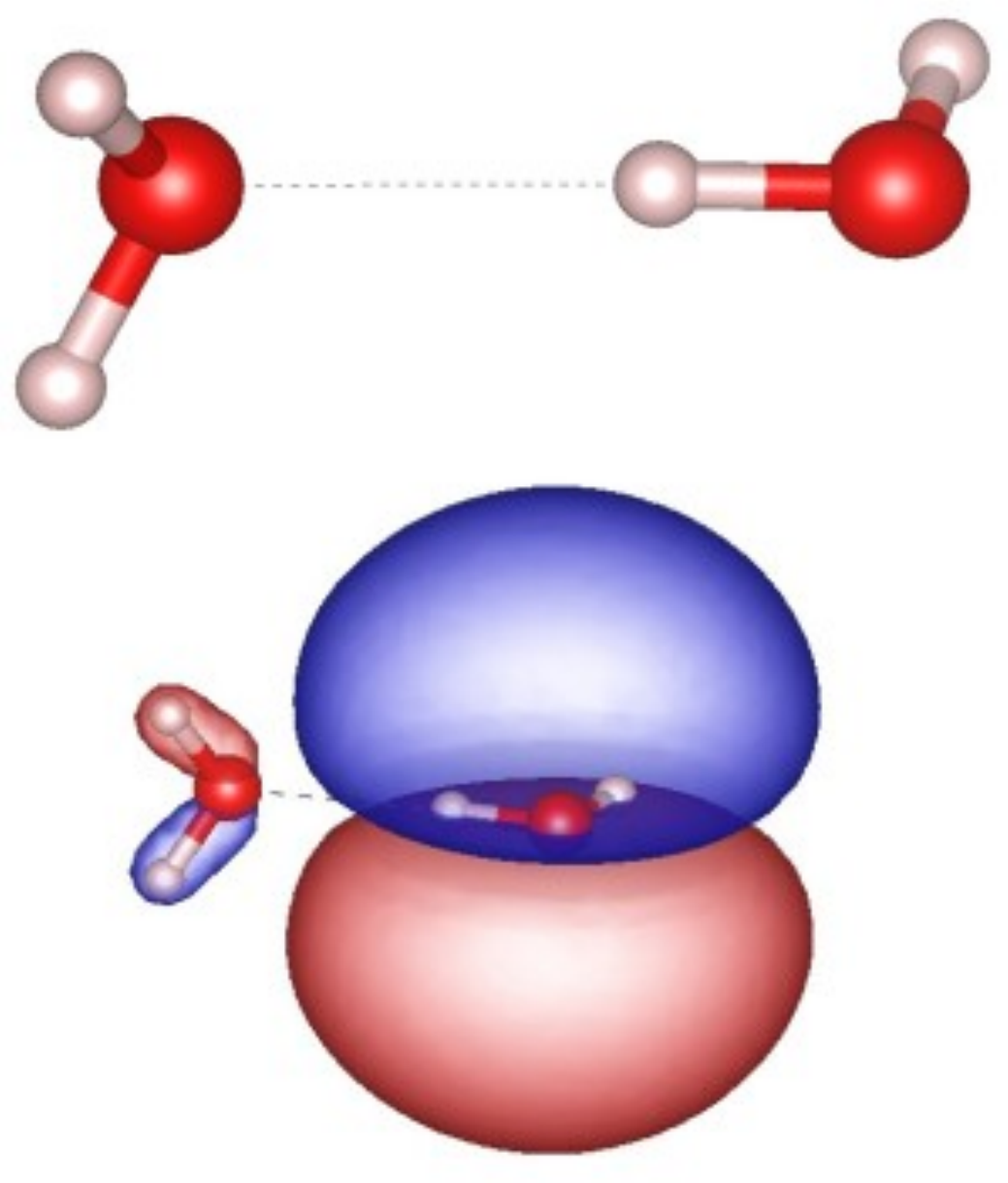}
			\caption{}
			\label{fig:h2o_N}
		\end{subfigure}	
		\begin{subfigure}[b]{0.3\textwidth}
			\includegraphics[width=1.3in]{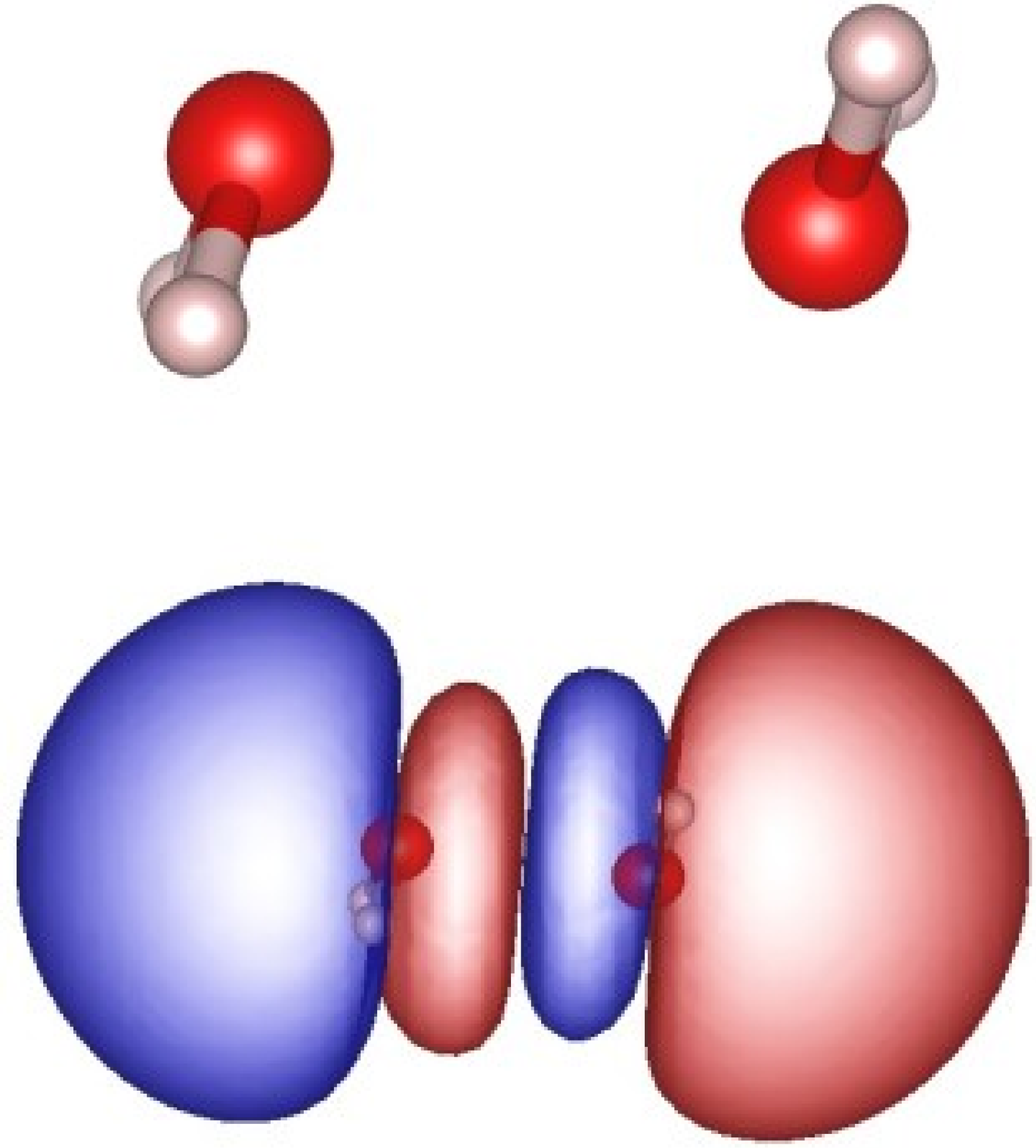}
			\caption{}
			\label{fig:h2o_HB}
		\end{subfigure}
		\begin{subfigure}[b]{0.3\textwidth}
			\includegraphics[width=1.2in]{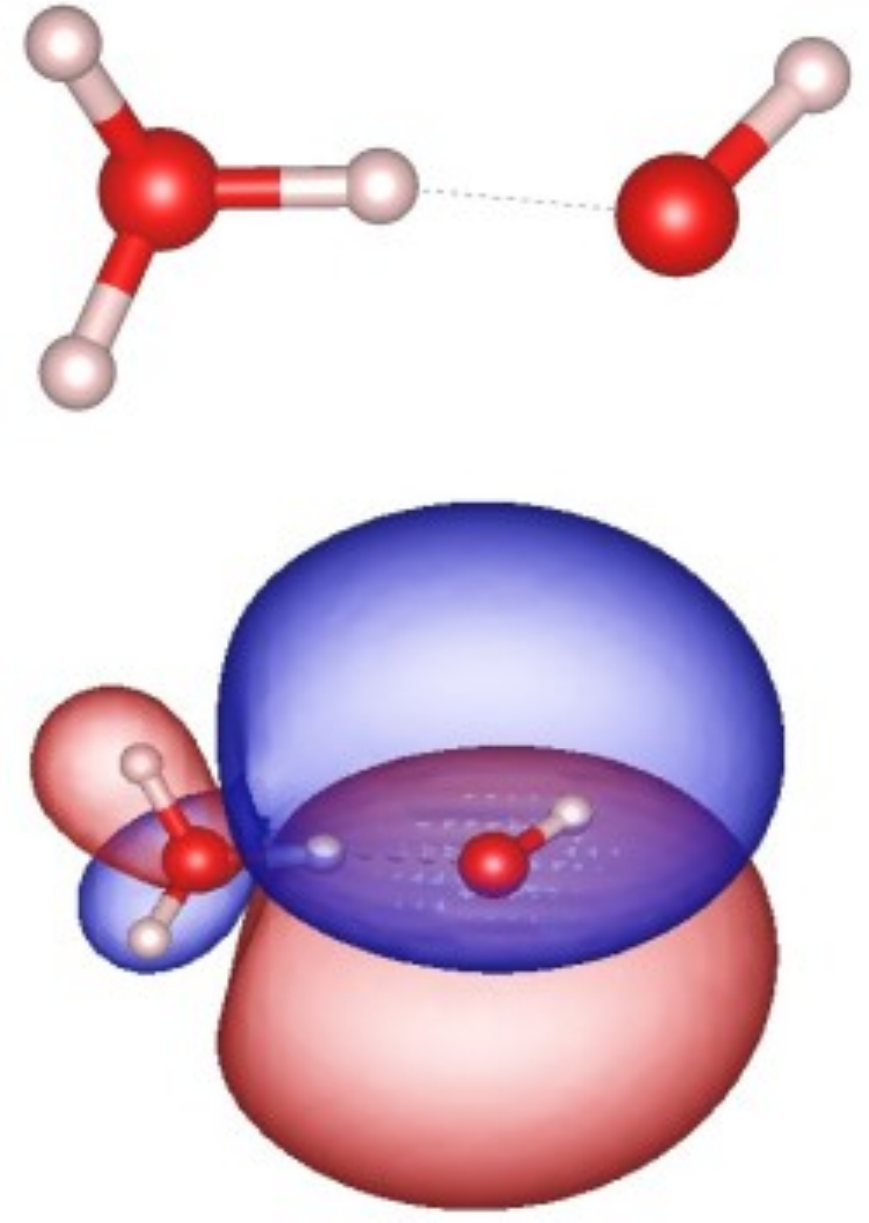}	
			\caption{}
			\label{fig:h2o_PT}
		\end{subfigure}
		\caption{Neutral water dimer (H$_2$O)$_2$ geometries together with their highest occupied molecular orbitals (HOMO) represented by the blue and red-shaded regions (isosurface value $= 0.01$ au$^{-3}$), given by PBE: (a) H-bonded, (b) Hemibonded, (c) Proton-Transferred.}
	\end{figure*}
	
	We choose small clusters of (H$_2$O)$_n$ ($n = 2-5$) as our model systems. 
	For the water dimer (H$_2$O)$_2$, we adopt the optimized (un-ionized) H-bonded geometry as shown in Fig. \ref{fig:h2o_N}, which corresponds to the energy minima of the neutral system \cite{doi:10.1021/ct300832f}.
	For $n = (3,4,5)$, all of the chosen (H$_2$O)$_n$ clusters exhibit open-framework linear chains comprising of unidirectional H-bonds.
	This ensures that the cooperative-strength of H-bonds in the system grows with the length of the chain ($n$) \cite{Guevara-Vela2016}.
	These structures do not correspond to a minima in the potential energy surface of (H$_2$O)$_n$ ($n > 2$) water clusters, which have a closed H-bond network, optimizing the formation of four H-bonds per water molecule \cite{Brini2017}.
	This emphasizes the importance of H-bond cooperativity effects in studying the non-adiabatic dynamics of ionized (H$_2$O)$_n^+$ clusters. 
	More importantly, this choice ensures that the photo-generated hole is always localized on the first molecule of the chain, which always forms a single donor H-bond.
	We first perform a Born-Oppenheimer molecular dynamics (BOMD) simulation for the neutral (H$_2$O)$_n$ ($n = 2-5$) system, wherein the nuclei evolve on a single potential energy surface and the electronic structure is solved self-consistently using the DFT module of NWChem package \cite{VALIEV20101477}. 
	We use a 6-31++G** basis set and the BHLYP hybrid functional. The adiabatic ground-state dynamics for each of the neutral (H$_2$O)$_n$ systems is performed at a temperature of 200K using a Langevin thermostat for 100 ps, with a time step of 0.5 fs.
	Uncorrelated snapshots are chosen from this single adiabatic trajectory (simulated for each of the $n = 2-5$ systems) at random time intervals, to be further used as initial geometries for the excited-state non-adiabatic simulations upon HOMO-ionization at $t = 0$. 
	All the structures derived from the adiabatic simulations maintain mostly linear one-dimensional H-bonds connecting each water to its neighbor(s). 
	
	\subsection{Non-adiabatic Molecular Dynamics}
	
	The geometries extracted from the BOMD trajectory are ionized by removing an electron from the HOMO of the system, i.e. setting the net charge of the system to $+1$. 
	We then perform rt-TDDFT-based Ehrenfest dynamics to simulate the excited-state dynamics, in which the electron dynamics is treated quantum mechanically, and the nuclei are classically propagated on a single mean-field surface given by an average over several electronic states.
	The TDDFT-based Ehrenfest dynamics of electrons and nuclei is formally given by:
	\begin{eqnarray}
	M_J\frac{\partial^2 \mathbf{R}_J}{\partial t^2} &=& 
	-\nabla_{\mathbf{R}_J}\widehat{V}_{nn}(\{\mathbf{R}_J(t)\})
	\nonumber\\
	&&-\int d^3r\rho(\mathbf{r},t)\nabla_{\mathbf{R}_J}\widehat{V}_{
		en}(\{\mathbf{R}_J(t)\}) ~,\\
	i\frac{\partial \phi_i (\mathbf{r},t)}{\partial t} &=& 
	\left(\frac{-\nabla^2}{2}+v_{KS}(\mathbf{r},t) 
	\right)\phi_i(\mathbf{r},t) ~,
	\end{eqnarray}
	where $\mathbf{R}_J$ denotes the nuclear coordinates and $\phi_i$ are the time-dependent Kohn-Sham orbitals such that the electronic density is given by: $\rho(\mathbf{r},t)=\sum_i\vert\phi_i(\mathbf{r},t)\vert^2$. 
	$\widehat{V}_{nn}$ and $\widehat{V}_{en}$ are the nuclear and electron-nuclei interaction terms respectively, 
	\begin{eqnarray*}
		\widehat{V}_{nn}&=&\frac{1}{2}\sum_{I\neq J}\frac{Z_I Z_J}{\vert \mathbf{R}_I(t) -\mathbf{R}_J(t)\vert} ~,\nonumber\\ 
		\widehat{V}_{en}&=&-\sum_I Z_I\int d^3r' \frac{\rho(\mathbf{r'},t)}{\vert \mathbf{r'} -\mathbf{R}_I(t)\vert} ~.
	\end{eqnarray*}
	The time-dependent Kohn-Sham potential $v_{KS}(\mathbf{r},t)$ is written as 
	\begin{equation}
	v_{KS}(\mathbf{r},t) = v_{\text{ext}}(\mathbf{R},\mathbf{r},t)+v_{\text{H}}[\rho](\mathbf{r},t)+v_{\text{xc}}[\rho](\mathbf{r},t) ~,
	\end{equation}
	where $v_{\text{ext}}(\mathbf{R},\mathbf{r},t)$ is the external potential due to moving nuclei, $v_{\text{H}}[\rho](\mathbf{r},t)=\int d^3r'\frac{\rho(\mathbf{r'},t)}{\vert \mathbf{r} -\mathbf{r'}\vert}$ gives the Hartree potential and $v_{\text{xc}}[\rho](\mathbf{r},t)$ is the XC potential (derived from the approximate E$_{\text{XC}}[\rho]$).

	The non-adiabatic simulations performed for multiple HOMO-ionized ($+1$) geometries give rise to an ensemble of Ehrenfest trajectories (ETs). 
	We sample these ETs to obtain the relevant time-scales and mechanisms of the proton-transfer reaction occurring in (H$_2$O)$_n^+$.
	The rt-TDDFT/Ehrenfest dynamics simulations are performed using the real-space grid code, Octopus-8.2 \cite{C5CP00351B}. 
	The optimal values of grid spacing and size of the simulation box are obtained from energy convergence tests on the systems. 
	We use a default spacing of 0.23 \AA\, and a spherical simulation box with a radius of 8 \AA\, for the dimer, 10 \AA\, for the trimer, and 15 \AA\, for all other water chains.
	The coupled electron-ion dynamics uses an enforced time-reversal symmetry (ETRS) propagator \cite{doi:10.1063/1.1774980} with a time step of 1.3 attoseconds (as). 
	This was determined to be the maximum time step that conserves the energy within a suitable range (0.7 eV over 20 fs), and the nuclear dynamics is similar to that obtained with smaller time steps (see supplemental Fig. \ref{fig:energydimer}, \ref{fig:distancedimer}).
	All the ETs are propagated for up to 200 fs in order to simulate explicit dissociation or molecular rearrangement in the excited system.
	
	\section{\label{sec:molecular_dynamics}Results and Discussion}
	
	\subsection{rt-TDDFT diabatic trajectories}
	
	\subsubsection{Dimer (H$_2$O)$_2^+$}
	An extensive study focusing on the dynamics of ionized water monomer and dimer was carried by Chalabala \textit{et al.} \cite{doi:10.1021/acs.jpca.8b01259}. 
	They compared the results from two non-adiabatic approaches, namely surface hopping and Ehrenfest dynamics, and highlighted the importance of using hybrid functionals in the latter method to obtain accurate simulation outcomes. 
	In this work, we consistently employ the Ehrenfest dynamics approach, but focus on a non-hybrid functional (PBE).
	We also confirm the results obtained by Chalabala \textit{et al.} \cite{doi:10.1021/acs.jpca.8b01259} for the smallest system, that is, a dimer cation

	The HOMO of the neutral (H$_2$O)$_2$ system is localized on the H-bond donor molecule in the optimized H-bonded configuration as shown in Fig. \ref{fig:h2o_N}.
	Ionizing such a configuration results in an unpaired electron, or equivalently, a hole in this ``occupied'' orbital.
	\begin{figure}
		\begin{subfigure}[b]{0.45\textwidth}
			\includegraphics[width=2.5in]{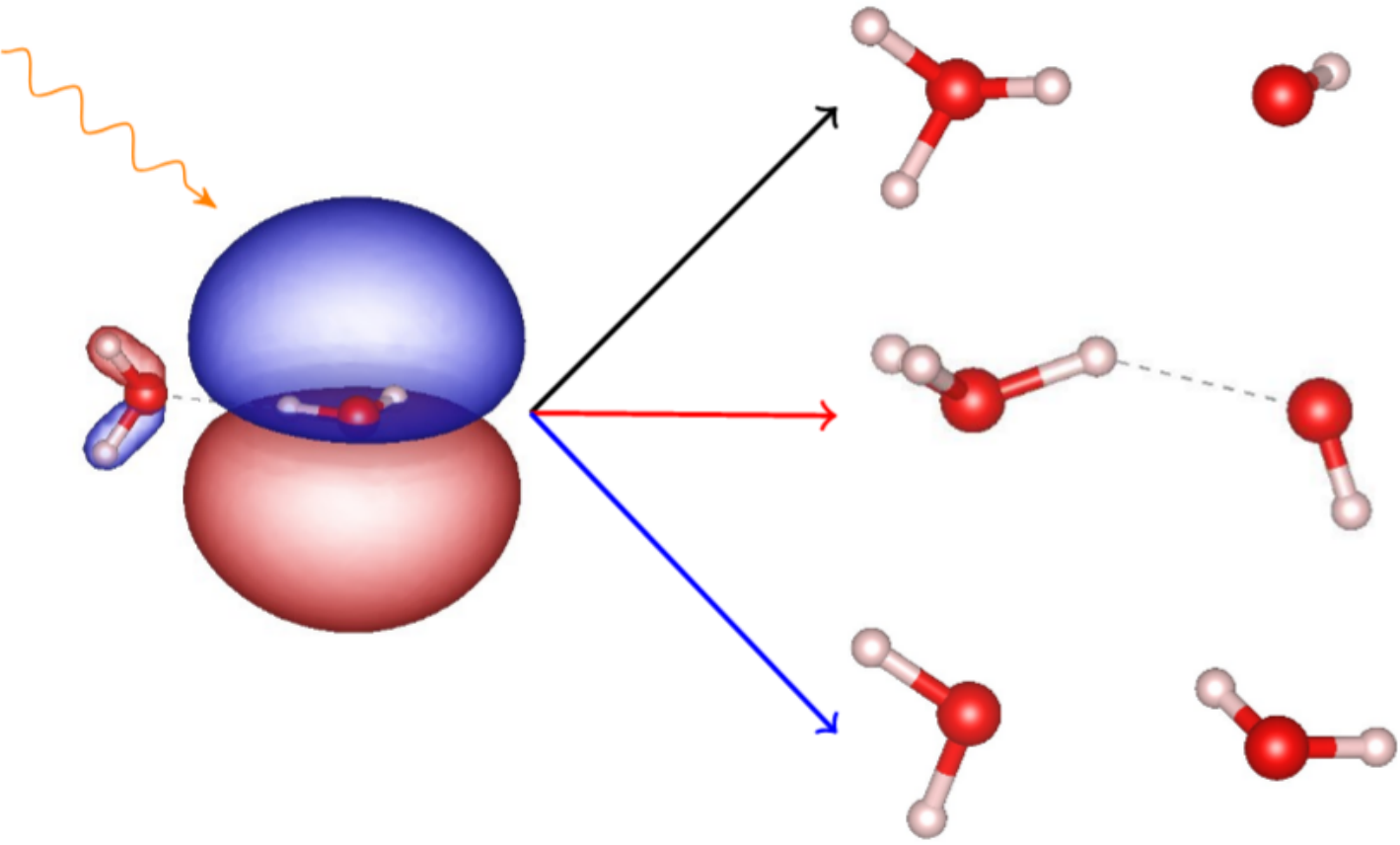}
			\caption{}
			\label{fig:fig1}
		\end{subfigure}
		\begin{subfigure}[b]{0.45\textwidth}
			\includegraphics[width=3.0in]{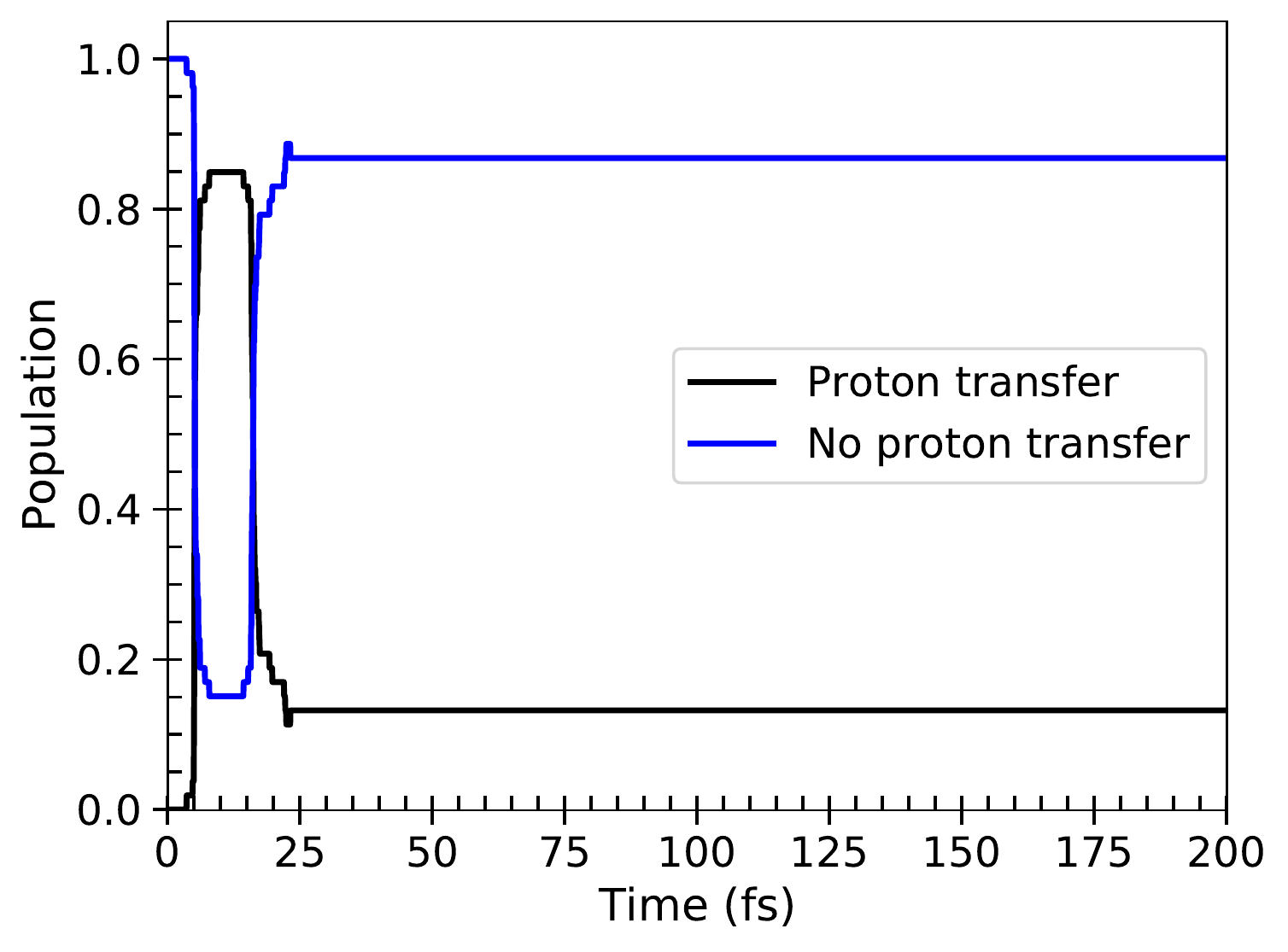}	
			\caption{}
			\label{fig:fig3}
		\end{subfigure}
		\caption{H-bonded water dimer, PBE: (a) Initial ($t=0^-$) structure and HOMO spin-density (blue and red-shaded regions, cutoff value = $0.01$ au$^{-3}$). (b) Population analysis for rt-TDDFT/PBE Ehrenfest dynamics of HOMO-ionized water dimer. Statistics obtained from 53 independent trajectories. The black and blue curves indicate two reaction channels -- proton transfer (H$_3$O$^+\ +\ ^{\bm{\cdot}}$OH) and no proton transfer or dimer dissociation (H$_2$O$^+$\ +\   H$_2$O) respectively.} 
	\end{figure}
	The time-evolution of a photoionized water dimer performed using rt-TDDFT/PBE produces one of the two reaction channels, shown in Fig. \ref{fig:fig3}.
	We choose to simplify the outcome channels into a binary -- proton transfer and no proton transfer -- classification.
	The proton transfer channel indicates that the proton from the ionized molecule is transferred to its immediate H-bonded neighbor. 
	This channel can be further divided into separate channels with bound and unbound molecules, see black and red arrows in Fig. \ref{fig:fig1}. However, we do not make this distinction in our trajectory-based statistics.
	In order to sample the population of these channels for the simulation period, we average over 53 individual Ehrenfest trajectories, see Fig. \ref{fig:fig3}. 
	The initial short-time Ehrenfest dynamics ($t< 15$ fs) is characterized by a rapid proton transfer seen in a majority (fraction of 0.85) of the simulated trajectories. 
	This is almost always followed by a proton bounce-back to the original donor oxygen, signaled by a decrease in the O$_d\cdots$H$^+$ bond length.
	This result confirms that in non-adiabatic Ehrenfest simulations, it is imperative to propagate the nuclei on the excited-state surface for longer times to conclusively determine the underlying mechanism and outcome of the simulated dynamics \cite{C6SC04378J}.

	The black curve in Fig. \ref{fig:fig3} shows the successful  proton-transfer trajectory statistics.
	A proton is considered as transferred when the H$^+\cdots$O$_a$ bond distance $d_{\text{O$_a$}\cdots \text{H}}< 1.5$ \AA, ($a=$ acceptor). 
	The second, no proton-transfer channel, results from a molecular rearrangement of the two water units with the net charge ($+1$) being shared by them. 
	In these trajectories, the two water monomers transition from a short-living proton-transferred geometry to a hemibonded-type configuration. 
	We observe a prevalence of this channel in the trajectory statistics using the semilocal PBE functional.
	At the end of the simulation period ($t=200$ fs), a proton-transferred structure is formed in only $13 \%$ of the initiated trajectories.
	On the other hand, using the BHLYP (hybrid) functional in Ehrenfest dynamics, Chalabala \textit{et al.} \cite{doi:10.1021/acs.jpca.8b01259} found $95 \%$ of their trajectories resulting in proton-transfer events.
	We associate this difference to the fact that PBE spuriously determines the hemibonded structure to be lower in energy $\Delta$E$_{\text{PT}\rightarrow \text{HB}} = -0.18$ eV (see Table \ref{tab:hbptbenergies}). 
	\subsubsection{Trimer (H$_2$O)$_3^+$}
	
	\begin{figure}
		\begin{subfigure}{0.45\textwidth}
			\includegraphics[width=1.5in]{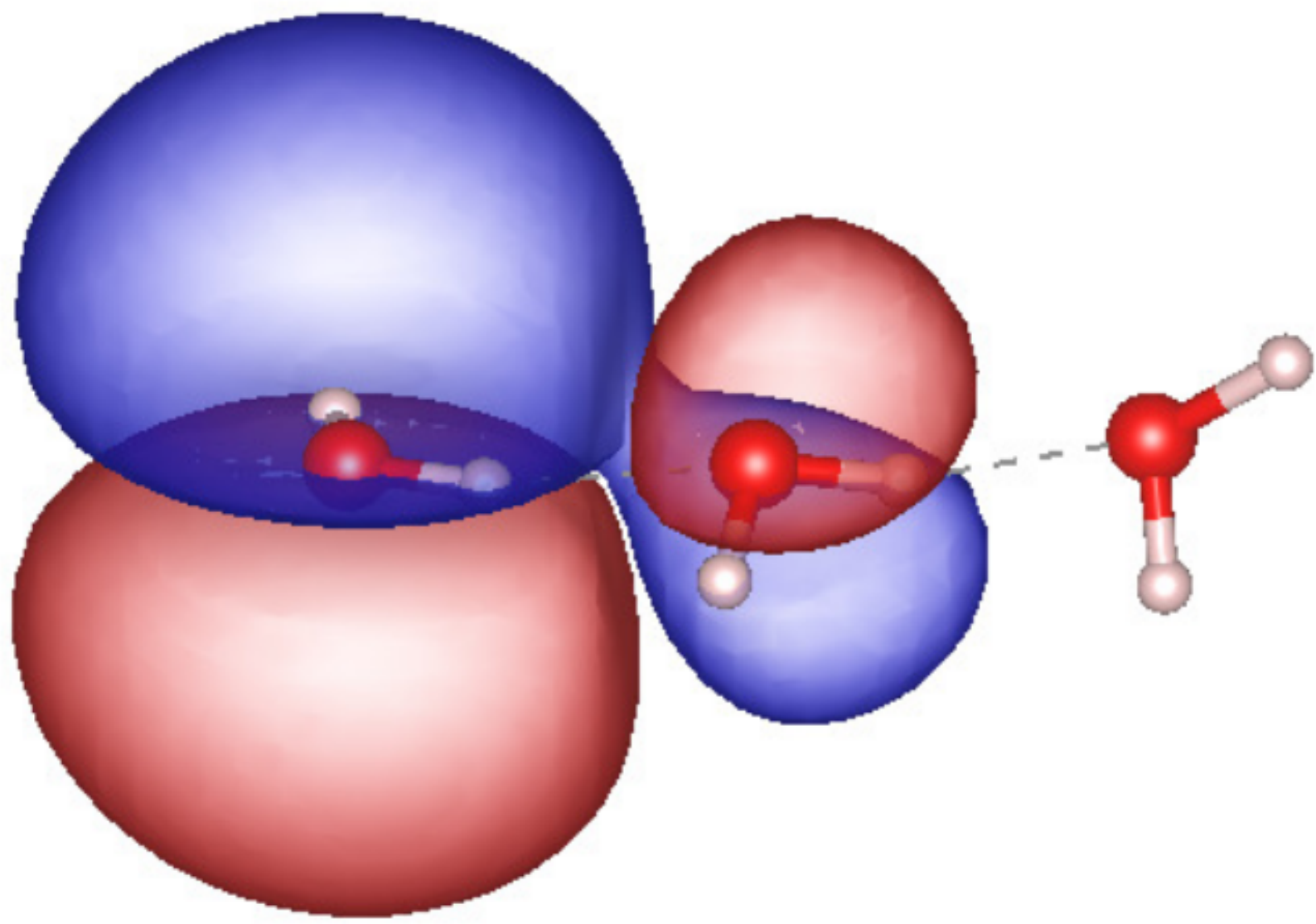}
			\caption{}
			\label{fig:fig5}
		\end{subfigure}
		\begin{subfigure}{0.45\textwidth}
			\includegraphics[width=3.0in]{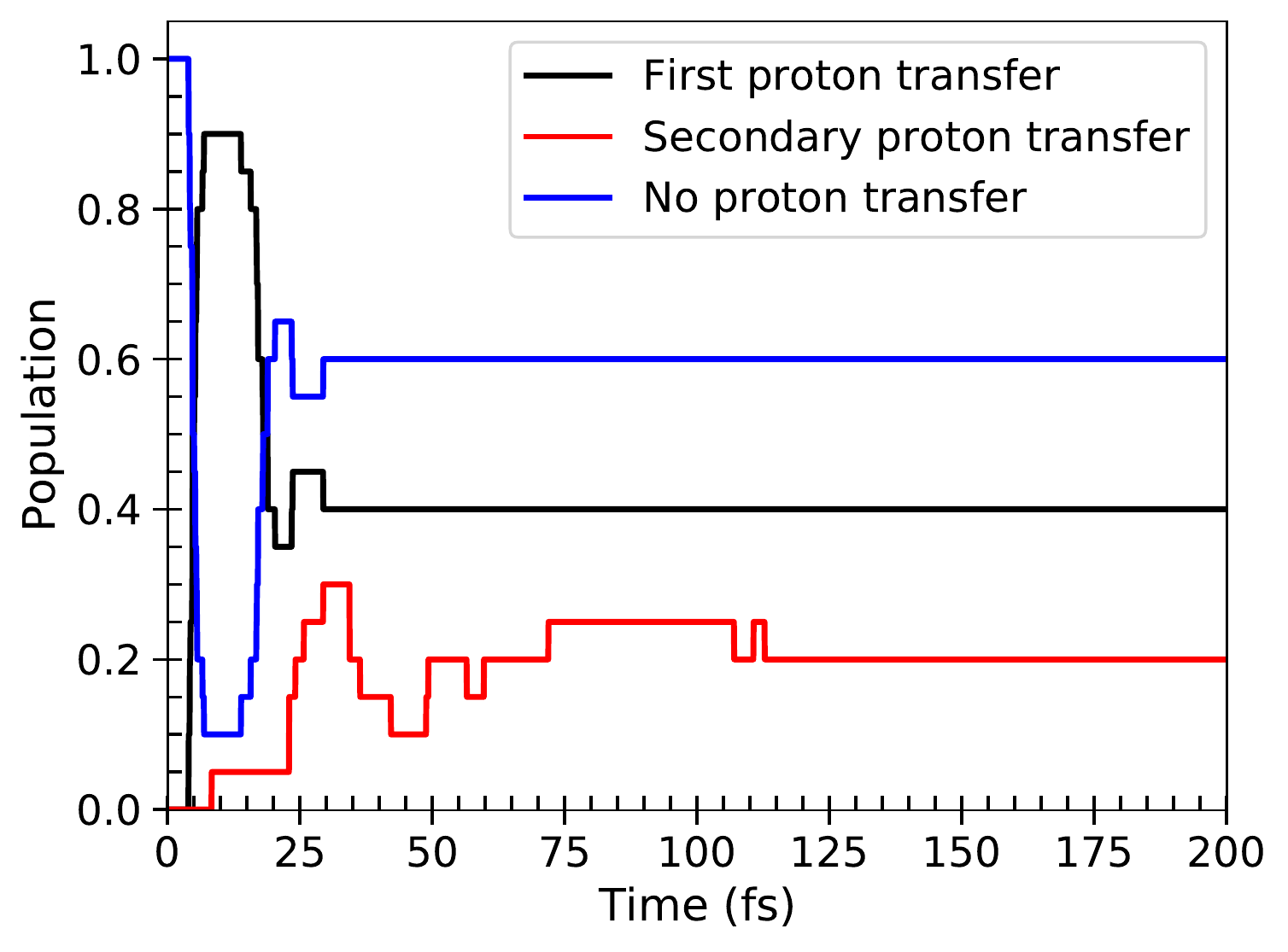}
			\caption{}
			\label{fig:fig6}
		\end{subfigure}
		\caption{H-bonded water trimer, PBE: (a) selected geometry and corresponding isosurface (cutoff value = $0.01$ au$^{-3}$) of the HOMO spin density (at $t = 0^-$). (b) Population analysis for rt-TDDFT Ehrenfest dynamics of the HOMO-ionized system. Statistics obtained from 20 independent trajectories. Note that the secondary proton transfer trajectories (red) are normalized with respect to the total number of trajectories and not with respect to their parent trajectories (black).}
	\end{figure}
	Fig. \ref{fig:fig6} shows the proton-transfer population statistics for chainlike structures of ionized water trimers. 
	These structures contain an extra H-bond (compared to the dimer), and the HOMO spin density in a corresponding neutral trimer chain is localized on the single H-donating oxygen, with some weight on the H-bond acceptor, see Fig. \ref{fig:fig5}.
	The trajectories supporting a proton-transfer suggest the formation of a long-lasting intermediate [H$_3$O$^+\cdots^{\bm{\cdot}}$OH] bonded-pair  before the fragments dissociate to give mobile hydroxyl radical and a reactive hydronium cation.
	For the 20 simulated trajectories, the first proton hop is observed to be relatively fast (within 10 fs in a fraction of 0.90 of the trajectories), and $40 \%$ of the trajectories result in a proton-transfer at the end of simulation period ($t=200$ fs). 
	This increase in proton-transfer population (in comparison to the dimer ion case) correlates with the energies in Table \ref{tab:hbptbenergies}. 
	The PBE energies listed in Table \ref{tab:hbptbenergies} indicate that $[E_{\text{HB}}-E_{\text{PT}}]_{\text{PBE}}<0$. 
	However, the magnitude of the energy difference, $|E_{\text{HB}}-E_{\text{PT}}|_{\text{PBE}}$, reduces in going from $n=2$ to 3, making the hemibonded geometry in trimer not as highly favorable as it is for dimer. 
	Moreover, the presence of another H-bonded water (right-most) in this chain makes it less likely for the central water to twist out of the H-bonded geometry  into a hemibonded-type one.

	In some trajectories ($20 \%$), a secondary proton-transfer from H$_3$O$^+$ to the nearby H$_2$O occurs. 
	The simulations reveal two key factors that influence such a process: the interatomic O$\cdots$O distances, and the time it takes for the initially nonparticipating H$_2$O to break away from the two interacting water-monomers compared to the first (primary) proton-transfer time. 
	If the first proton-transfer takes too long to complete due to excessive proton-rattling between the donor and acceptor oxygens, the nonparticipating H$_2$O is driven away by the electronic forces acting on the system, and the likelihood of a secondary proton transfer in the trimer chain is severely reduced.
	Since a secondary proton transfer event eliminates the probability of return of the proton to its original donor ($^{\bm{\cdot}}$OH), the bump seen in the secondary proton transfer curve (Fig. \ref{fig:fig6}) at shorter times ($t \sim 30$ fs) contributes toward an increase in the overall ``first" P(PT).

	\subsubsection{Tetramer (H$_2$O)$_4^+$} 
	
	\begin{figure}
		\begin{subfigure}{0.45\textwidth}
			\includegraphics[width=1.9in]{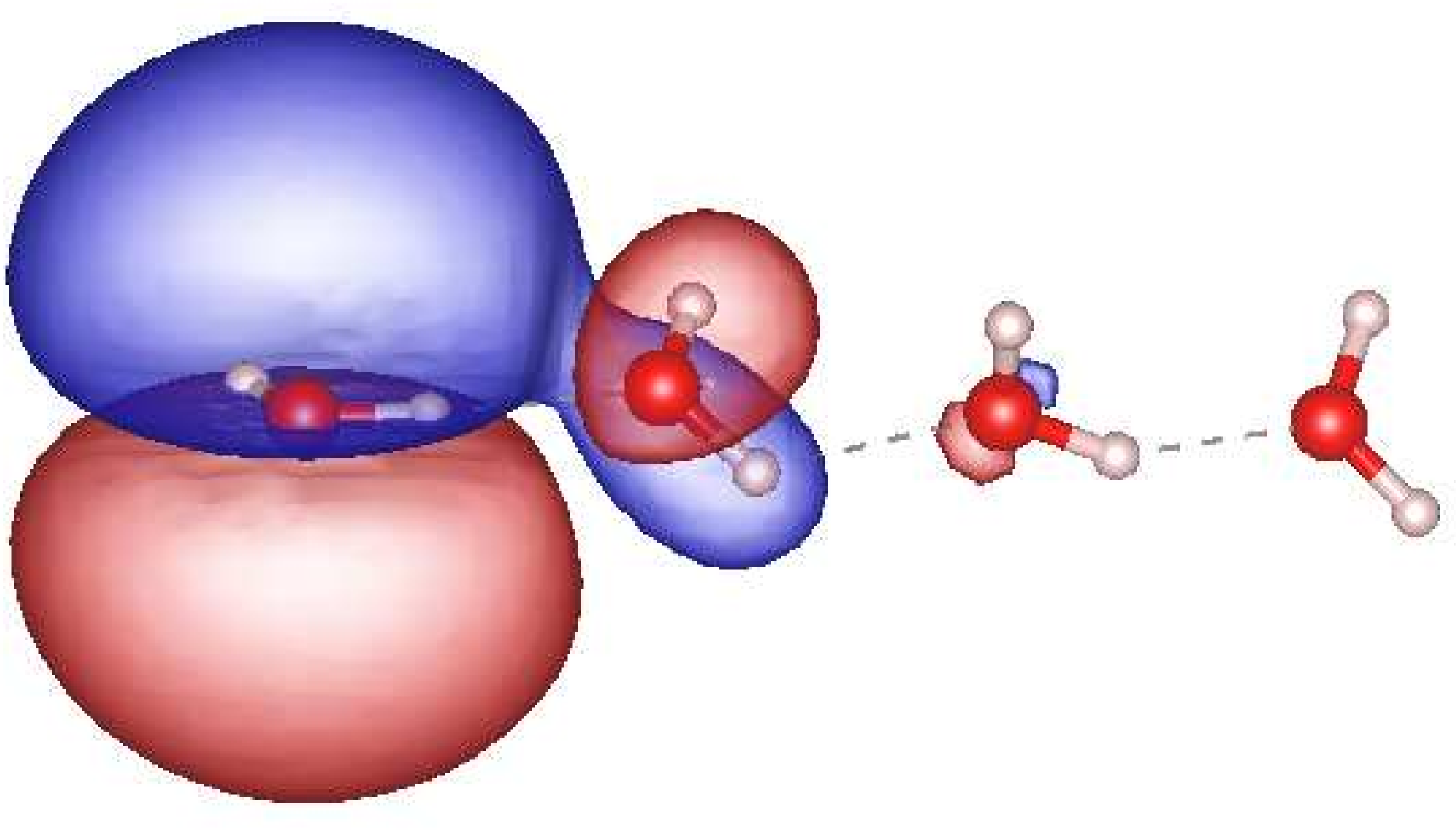}
			\caption{}
			\label{fig:fig8}
		\end{subfigure}
		\begin{subfigure}{0.45\textwidth}
			\includegraphics[width=3.0in]{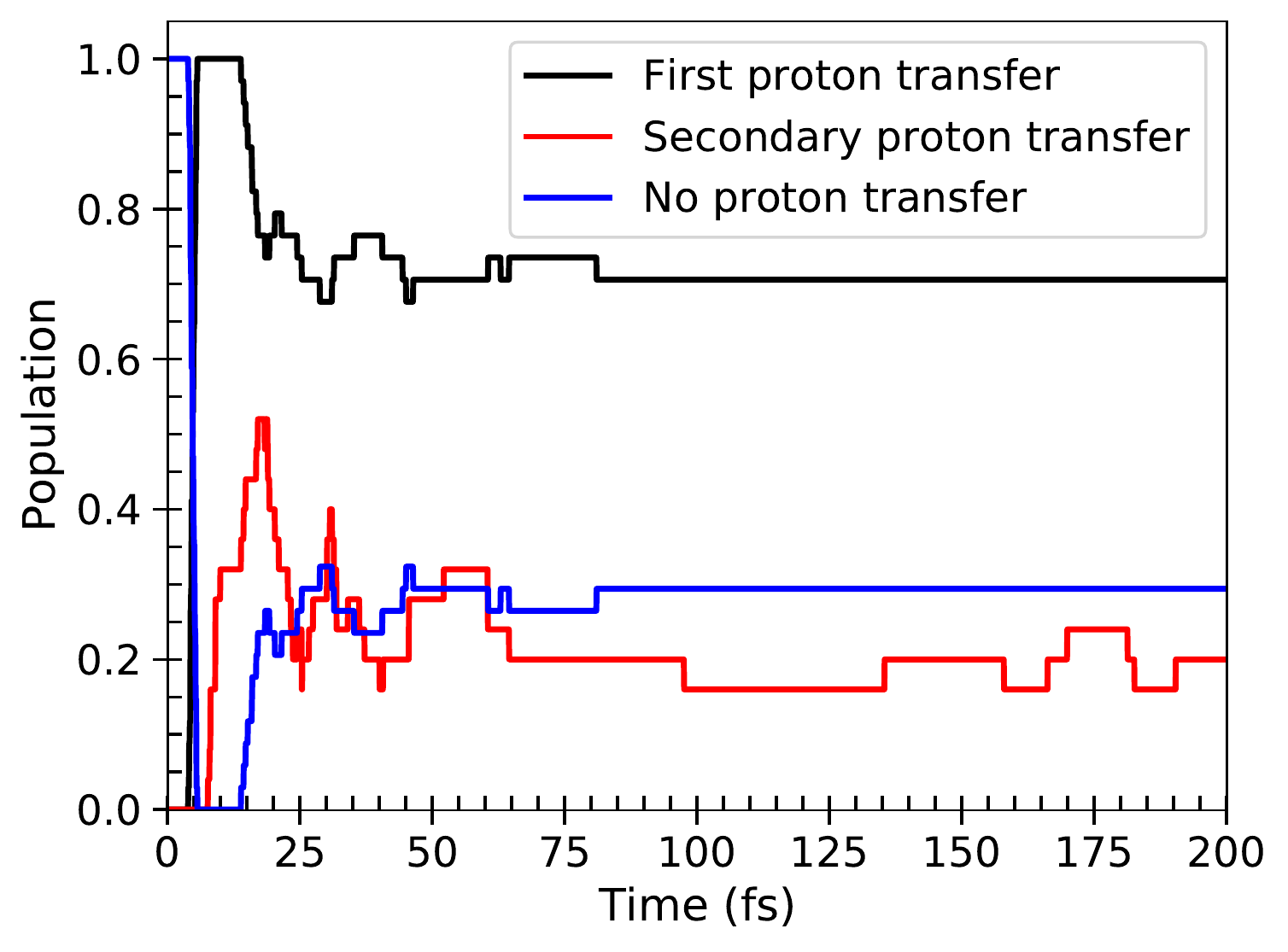}
			\caption{}
			\label{fig:fig9}
		\end{subfigure}
		\caption{H-bonded water tetramer, PBE: (a) selected geometry and corresponding isosurface (cutoff value = $0.01$ au$^{-3}$) of the HOMO spin density (at $t = 0^-$). (b) Population analysis for rt-TDDFT Ehrenfest dynamics of the HOMO-ionized system. Statistics obtained from 34 distinct trajectories.}
	\end{figure}
	As in the previous cases, in a neutral H-bonded water tetramer linear chain, the HOMO is primarily localized on the first molecule of the chain, which forms only one donor H-bond but no acceptor H-bonds (see Fig. \ref{fig:fig8}).
	Unlike the $n = 2, 3$ cases, all the tetramer ($n = 4$) trajectories exhibit a first proton-transfer step followed by fewer bounce-backs (only in $0.30$ of the 34 simulated ETs) to the original O-donor, as shown in Fig. \ref{fig:fig9}. 
	The fluctuations in proton-transfer population last up to 75-100 fs, after which a stable proton-transfer reaction statistics is obtained. 
	A successful proton-transfer  is observed for 70\% of the trajectories. 
	This increase correlates with the results in Table \ref{tab:hbptbenergies}, which shows how the energy difference between hemibonded and proton-transferred structures decreases with increasing chain lengths ($n$) in PBE. 
	That is, the self interaction error, while still present, becomes smaller.

	After the primary proton transfer reaction, the H$_{3}$O$^{+}$ cation may lose its proton to the neighboring H$_{2}$O (secondary proton transfer).
	This phenomenon can continue to propagate down the chain, depending on two factors: the length of H-bonded chain ($n$), and the time it takes for the linear order of H-bonds to disappear to form a gas phase, which we refer to as the H-bond chain lifetime, `$\tau$'. 
	For a particular system (and trajectory), $\tau$ serves as a cut-off after which the proton ceases to be pushed any further down the ``chain'' and maintains its position on the latest water unit. 
	In the case of a water dimer ion, frequent proton-hops back and forth (rattling events) between the donor and acceptor units result in a prolonged lifetime $\tau$. 
	At the same time, longer water clusters allow multiple proton transfers from one unit to the next in the chain, thereby also extending the lifetimes. 
	In general, a faster first proton transfer shortens the lifetime. 
	Additionally, the first and secondary proton-transfer events occur earlier with increasing system size $n$, see Table \ref{tab:lifetimes} for a comparison of all relevant time scales. 
	Therefore, the cooperative H-bonds in the ionized water chains induce a faster proton dynamics.
	
	\begin{table}[b]
		\caption{\label{tab:lifetimes}%
			First and secondary proton transfer times along with the H-bond chain lifetimes for a representative trajectory of (H$_2$O)$_n^+, n = (2-5)$.
		}
		\begin{ruledtabular}
			\begin{tabular}{cccc}
				&\multicolumn{1}{c}{First}&\multicolumn{1}{c}{Secondary}&\multicolumn{1}{c}{$\tau$}\\
				{n}&{proton transfer [fs]}&{proton transfer [fs]}&{[fs]}\\
				\colrule
				2 & 9.87 & - & 23.02 \\
				3 & 6.58 & 23.68 & 37.50 \\
				4 & 4.93 & 11.18 & 32.89 \\ 
				5 & 3.29 & 5.26 & 29.60 \\
			\end{tabular}
		\end{ruledtabular}
	\end{table}
	
	\subsubsection{Pentamer (H$_2$O)$_5^+$}
\begin{figure}
		\begin{subfigure}{0.45\textwidth}
			\includegraphics[width=2.2in]{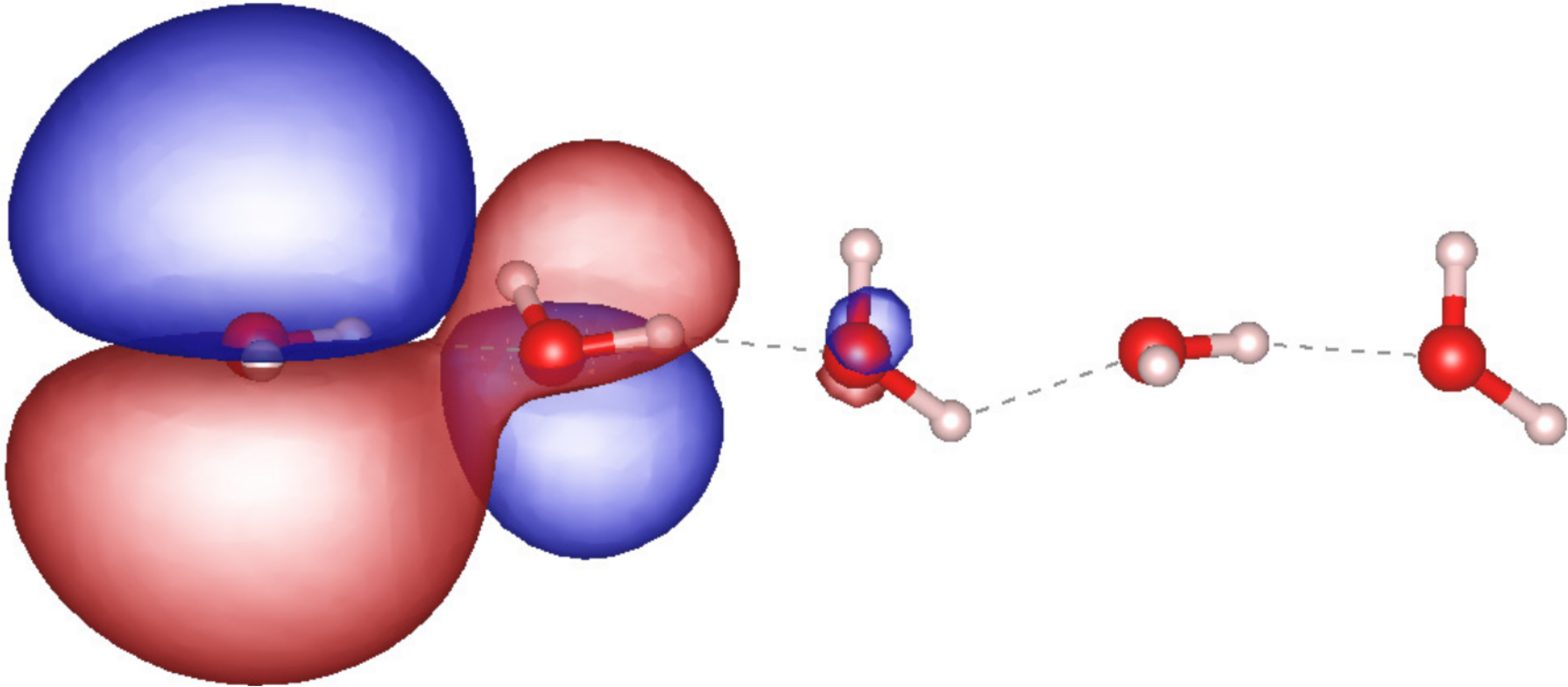}
			\caption{}
			\label{fig:fig11}
		\end{subfigure}
		\begin{subfigure}{0.45\textwidth}
			\includegraphics[width=3.0in]{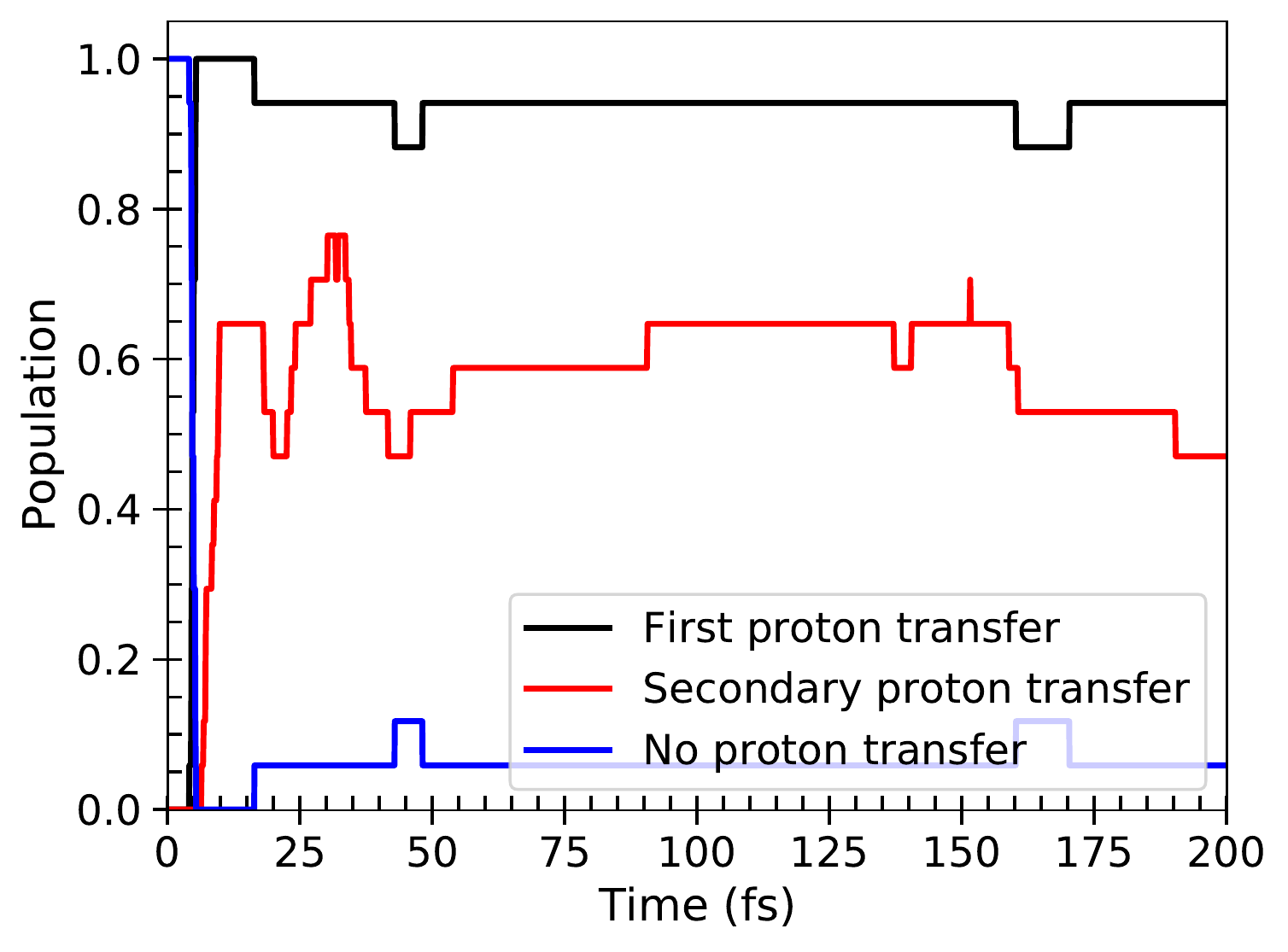}
			\caption{}
			\label{fig:fig12}
		\end{subfigure}
		\caption{H-bonded water pentamer, PBE: (a) selected geometry and corresponding isosurface (cutoff value = $0.01$ au$^{-3}$) of the HOMO spin density (at $t = 0^-$). (b) Population analysis for rt-TDDFT Ehrenfest dynamics of the HOMO-ionized system. Statistics obtained from 20 distinct trajectories.}
	\end{figure}

	In the pentamer chain, a fast first proton-transfer step is recorded within an average simulation time of 6.5 fs (for 20 simulated trajectories). 
	This is immediately followed by a secondary proton-transfer seen in most trajectories.
	94\% of the initiated (20) ETs predict proton-transfer reaction at the end of the simulation time ($t = 200$ fs) as depicted in Fig. \ref{fig:fig12}.
	Thus, rt-TDDFT/PBE gives a very high proton-transfer probability for an ionized pentamer chain. 
	The proton-transfer mechanism is governed by the formation of a bonded H$_3$O$^+$ ion $-$ $^{\bm{\cdot}}$OH radical contact pair at fixed O$_d-$O$_a$ distance.
	Subsequently, the two fragments separate after the occurrence of  another proton hop from H$_3$O$^+$ to the neighboring H$_2$O in the chain. 
	An important characteristic of this coupled electron-nuclear dynamics is that the separation of H$_3$O$^+-$ $^{\bm{\cdot}}$OH is fundamentally driven by the downhill electrostatic potential for the proton to move along the H-bonded water chain.

	With 94\% of the (H$_2$O)$_5^+$ ETs exhibiting a proton-transfer reaction, we expect the ionized pentamer chain ($n=5$) to be the PBE saturation limit as far as enhanced proton-transfer dynamics due to cooperative H-bonding interaction in water is concerned.
	This implies that for these sizes of chains, PBE and hybrid functionals should yield very similar results.
	For $n>5$, the (H$_2$O)$_n^+$ chains are likely to get divided into smaller sub-chains as the nuclei evolve in time, or they would prefer to exist as bifurcated (branched) structures -- both of which can be treated as a composite of the four cases ($n=2-5$) discussed in this work.
	
\begin{figure}
		\begin{subfigure}{0.45\textwidth}
			\includegraphics[width=2.0in]{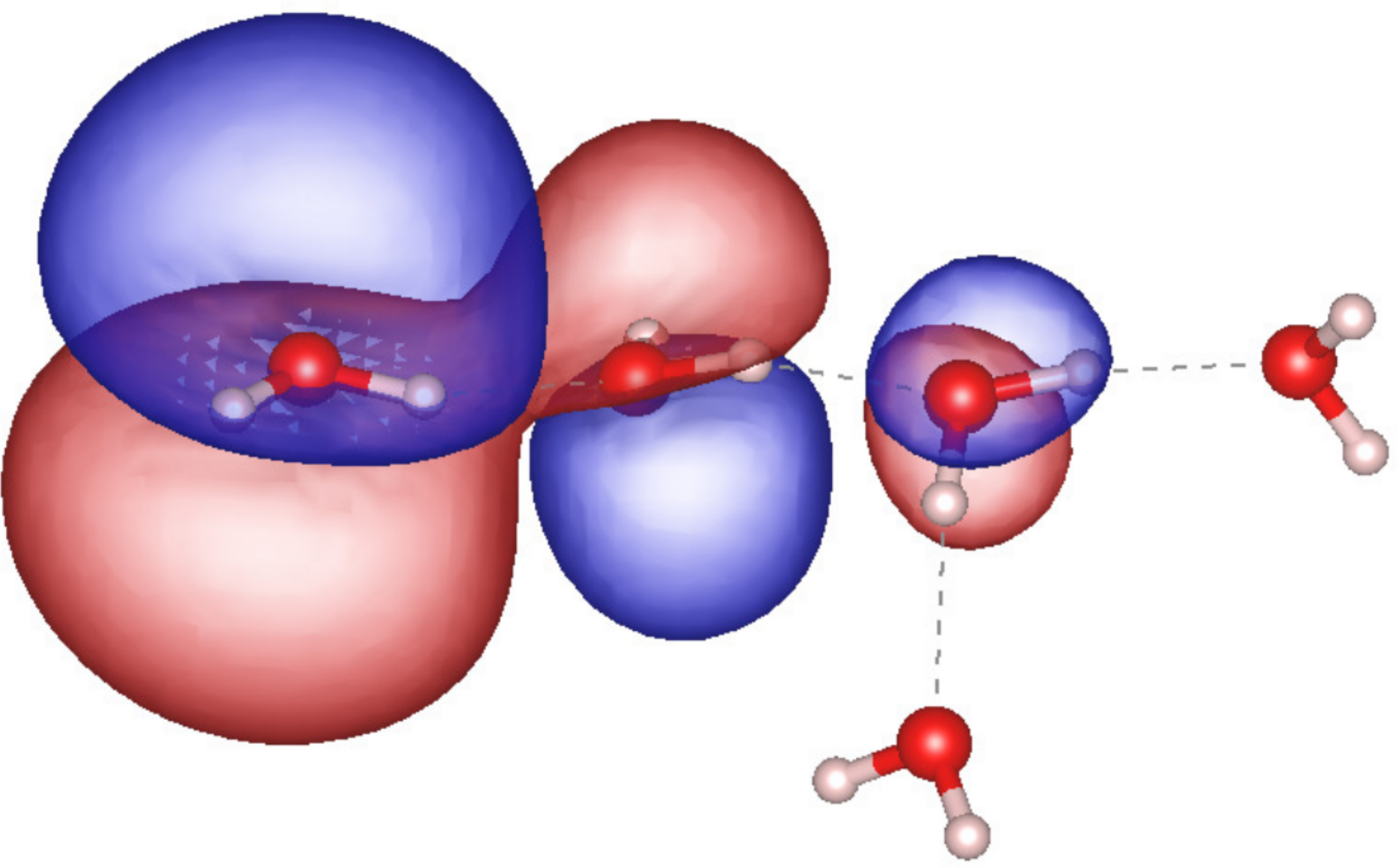}
			\caption{}
			\label{fig:fig14}
		\end{subfigure}
		\begin{subfigure}{0.45\textwidth}
			\includegraphics[width=3.0in]{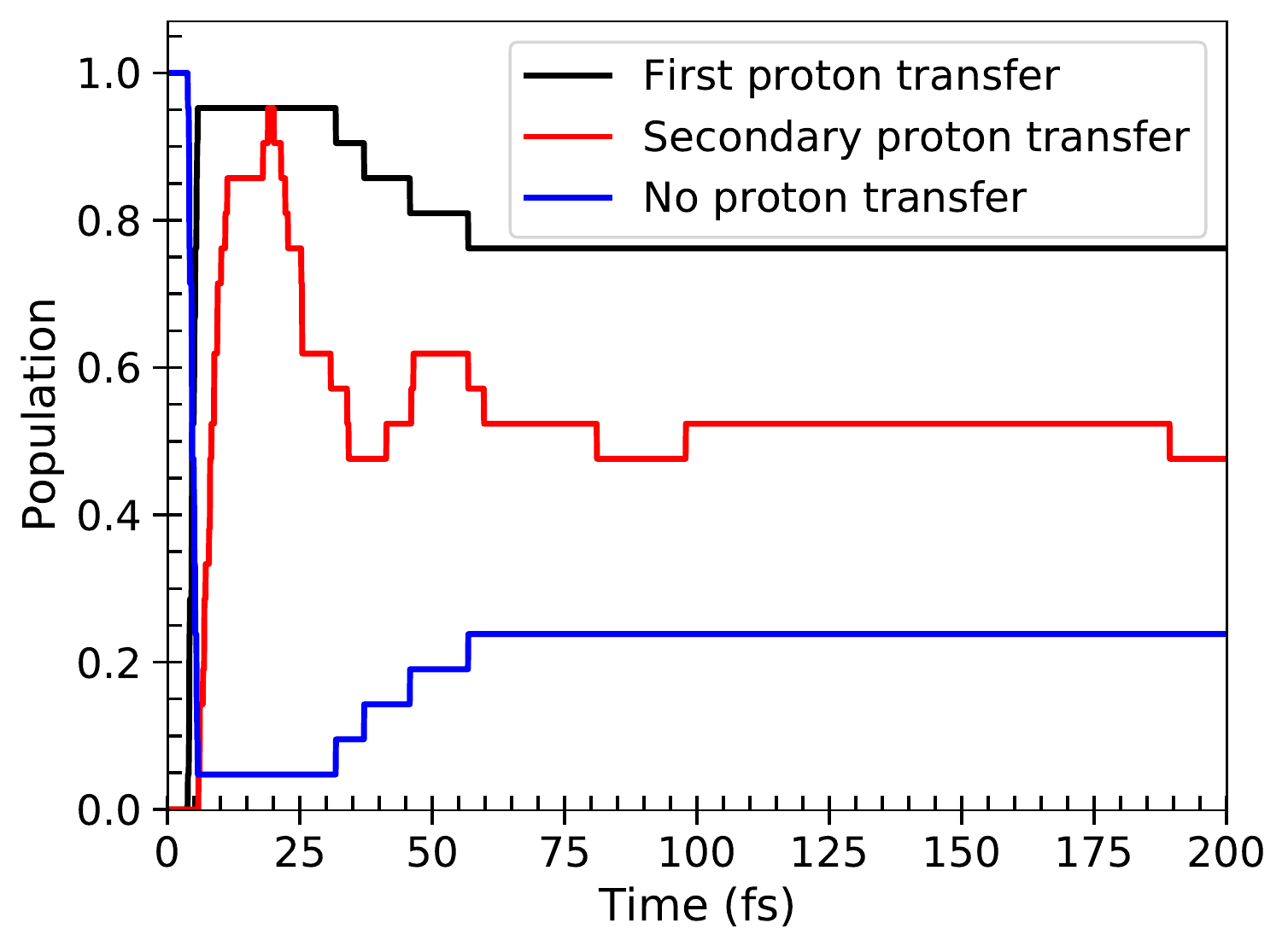}
			\caption{}
			\label{fig:fig15}
		\end{subfigure}
		\caption{H-bonded branched pentamer, PBE: (a) representative geometry and corresponding isosurface (cutoff value = $0.01$ au$^{-3}$) of the HOMO spin density (at $t=0^-$). (b) Population analysis for rt-TDDFT Ehrenfest dynamics of the HOMO-ionized system, (H$_2$O)-(H$_2$O)$_4^+$. Statistics obtained from 20 independent trajectories.}
	\end{figure}

	We have also considered the proton-transfer dynamics in a branched-pentamer chain.
	In this case, the alignment of H-bonds is no longer unidirectional, as a bifurcation allows one of the water units to form three H-bonds with its neighbors (instead of utmost two H-bonds per water exclusively studied so far), see Fig. \ref{fig:fig14}. 
	76\% of the 20 trajectories studied indicate proton-transfer over the simulation period, as shown in Fig. \ref{fig:fig15} . 
	This falls midway between the tetramer- and pentamer- Ehrenfest statistics, which is not surprising because the geometric arrangement can be viewed as either a tetramer chain with an extra water, or a pentamer chain with a (non-linear) displaced water.
	A distinctive feature of this type of chain is that all the trajectories supporting a first proton transfer also show a transient secondary proton-transfer event, see Fig. \ref{fig:fig15} at $t\sim$ 25 fs.
	At the end of the simulation window $\sim 50\%$ of the simulated trajectories result in a secondary proton-transfer , similar to the case of linear (H$_2$O)$_5^+$.
	This is because H$_3$O$^+$ formed after a secondary proton-transfer is stabilized by donating two H-bonds.
	This final structure is also found to occur in the linear pentamer chain.
	
	\subsection{\label{sec:hole_dynamics}Role of hole localization and dynamics}
	
	Our results indicate that for PBE, the P(PT) in HOMO-ionized H-bonded water chains increases with the length of the chain.
	We also know \cite{doi:10.1021/acs.jpca.8b01259} that hybrid functionals predict a P(PT) for the dimer similar to what PBE predicts for the pentamer chain.
	Our study has pointed to the SIE associated with an overestimation for hemibonded-type structures in order to explain the differences between the results for hybrid and non-hybrid functionals.
	However this is not enough to understand all the results. 
	In particular, the role that the localization of the photoionized hole plays on the P(PT) also needs to be considered.
	This is also important to understand to what point non-adiabatic simulations are necessary to accurately describe the rate of proton transfer upon single-ionization. 
	In their study of the ionized water dimer, Chalabala \textit{et al.}  \cite{doi:10.1021/acs.jpca.8b01259} showed that the rate of proton transfer is different in BOMD and rt-TDDFT simulations, both for hybrid and GGA functionals. 
	Here we evaluate this in an (H$_2$O)$_4^+$ chain, comparing PBE and PBE0.
	As a proof of concept, we choose $n=4$ for BOMD among all lengths considered in this study.
	
	\subsubsection{Photo-hole localization in adiabatic molecular dynamics}
	\begin{figure}
		\includegraphics[width=3.5in]{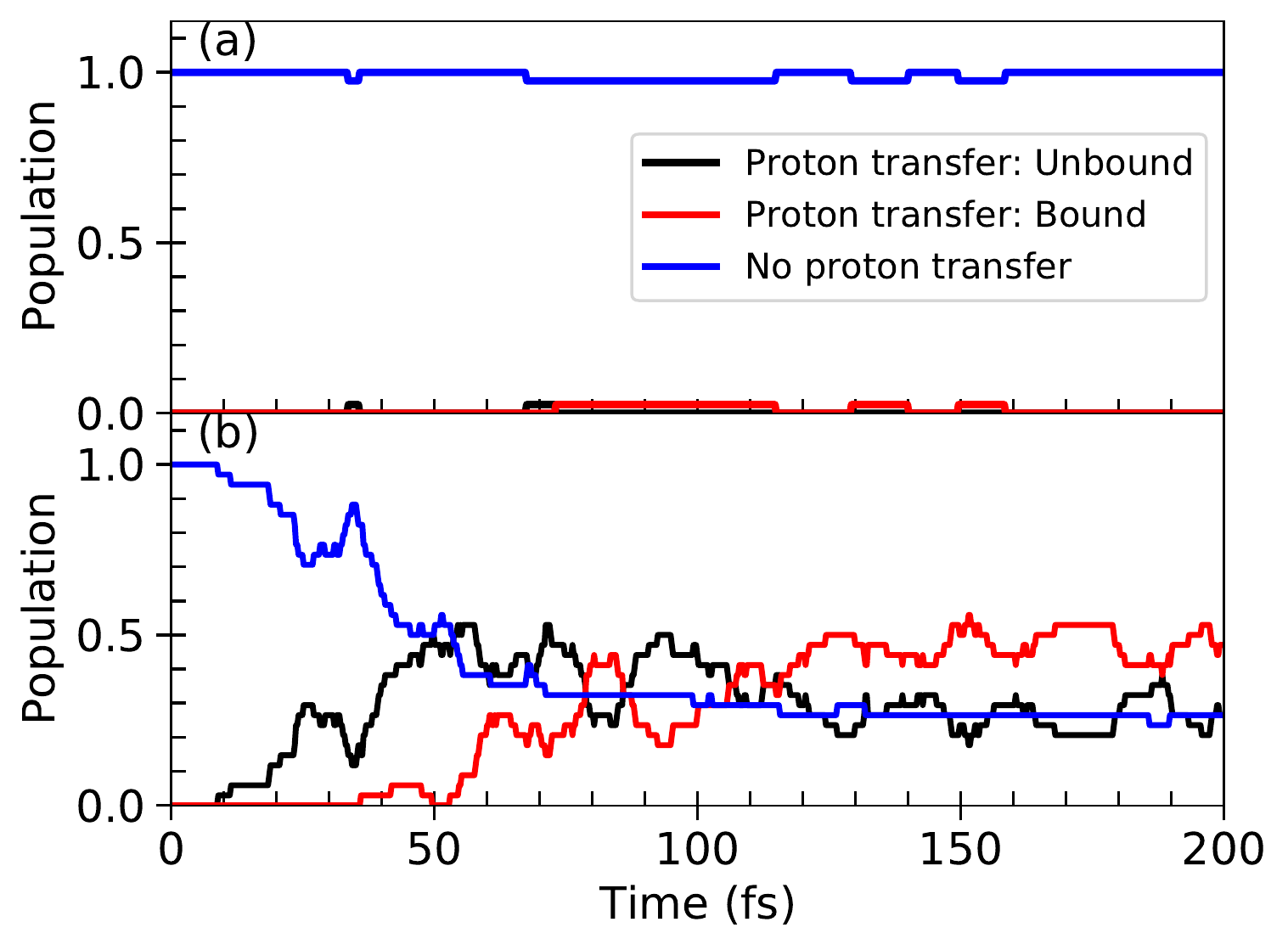}
		\caption{Adiabatic dynamics of HOMO-ionized water tetramer chain (H$_2$O)$_4^+$: (a)BOMD/PBE, (b) BOMD/PBE0.}
		\label{fig:fig13}
	\end{figure}
	
	\begin{figure*}
		\includegraphics[width=7.in]{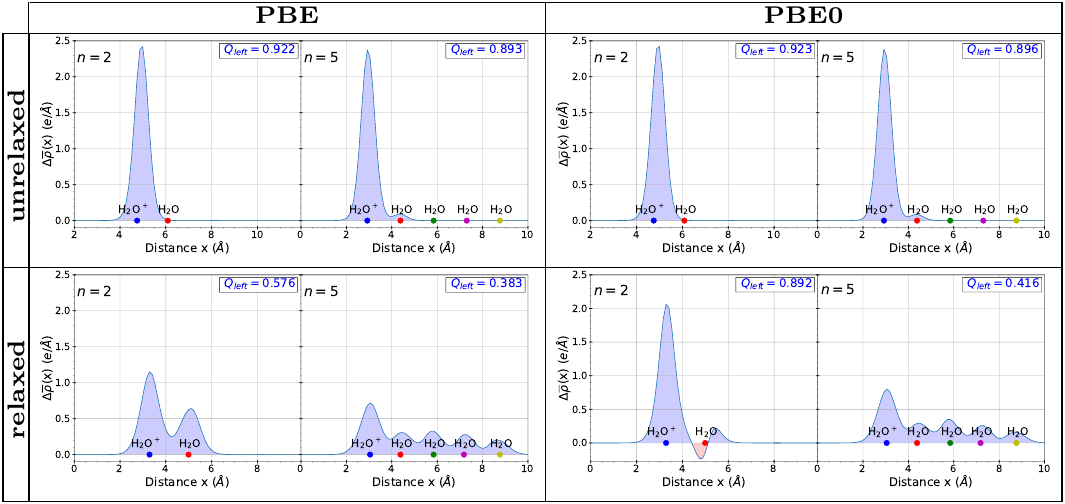}
		\caption{Hole densities given by the electronic density differences between ground-state and ionized (+1) configurations at $t=0$, shown here for ionized dimer and pentamer ($n=2,5$) water structures. \textbf{Unrelaxed} and \textbf{relaxed} refer to the type of wavefunction used to describe the ionized system.}
		\label{fig:holespbebhlyp2}
	\end{figure*}
	
	\begin{figure*}
		\includegraphics[width=5.8in]{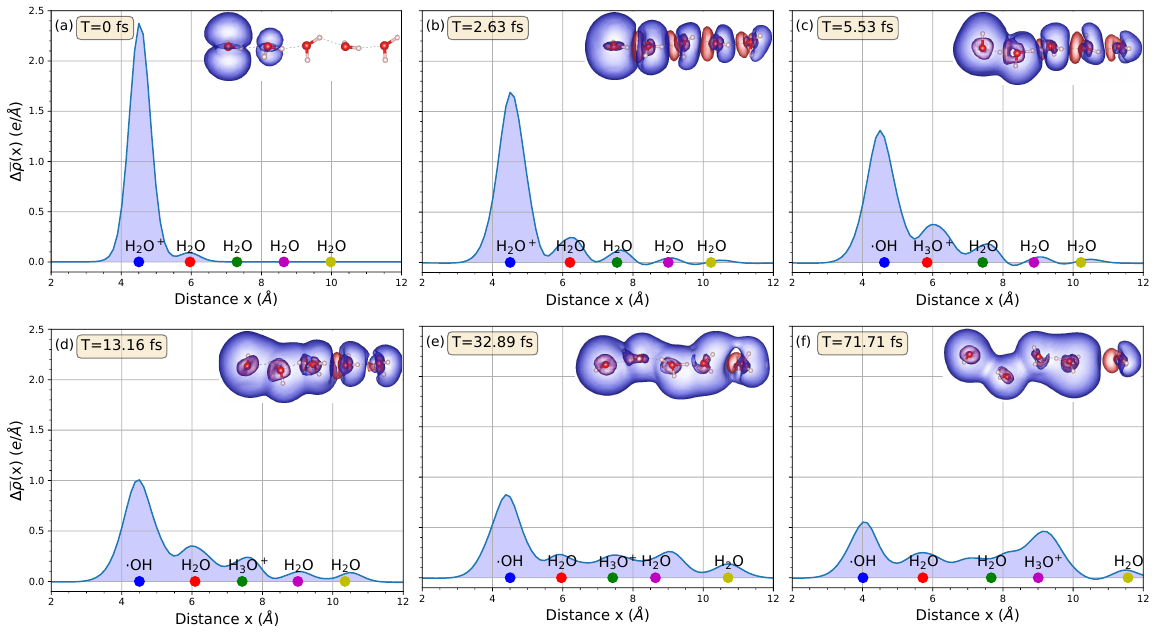}
		\caption{Snapshots of the electronic density differences between ground-state and ionized (+1) configurations obtained at various times showing the time-evolution of hole density for a single (H$_2$O)$_5^+$ Ehrenfest trajectory (rt-TDDFT/PBE). $\mathrm{T}= 0$ indicates the hole created at the time of ionization of the water chain.}
		\label{fig:hole_densities}
	\end{figure*}
	
	The BOMD simulations were performed using the NWChem package \cite{VALIEV20101477} with a time-step of 0.25 fs for the propagation of nuclei.
	BOMD constrains the evolution of the system on a purely adiabatic potential energy surface (PES), and does not time-evolve the electronic states, which therefore,  instantaneously adapt to the moving nuclei, with a parametric dependence on the nuclear coordinates. 
	As shown in Fig. \ref{fig:fig13} for (H$_2$O)$_4^+$, BOMD/PBE dynamics does not display any proton-transfer over 200 fs, whereas BOMD/PBE0 simulations show the formation of an unbound [H$_3$O$^+\cdots^{\bm{\cdot}}$OH] for 30\% of the 34 trajectories.
	The early-time features of the BOMD obtained at PBE0 are very different from the rt-TDDFT-based proton-transfer dynamics that we have seen so far. 
	In particular, there is a slow albeit gradual transfer of proton and fewer reverse bounces of the transferred proton. 
	This may well be a property of the \textit{adiabatic} propagation of the nuclei as there are no fluctuating electronic forces governing the motion of the proton.
	In principle, a comparison of the initial behavior of adiabatic and non-adiabatic dynamics highlights the mean-field averaging effect of the PESs inherent in the Ehrenfest approach \cite{doi:10.1002/cphc.201200941, Miyamoto2015, doi:10.1002/wcms.1417, doi:10.1063/1.2008258}.

	In the case of \textit{adiabatic} BOMD, PBE does not predict any proton-transfer (Fig. \ref{fig:fig13}(a)), while in PBE0 more than 50\% of the trajectories result in a proton transfer (Fig. \ref{fig:fig13}(b)). 
	We have shown in the previous section that in non-adiabatic coupled electron-nuclear dynamics (Fig. \ref{fig:fig9}), PBE predicts a proton-transfer rate of $\sim$ 70\% for the same system.
	The reason for such a discrepancy between adiabatic PBE and PBE0 on the one hand, and between adiabatic and diabatic PBE results on the other, is attributed to the nature of the wavefunction that the system is initialized in.
	In non-adiabatic Ehrenfest dynamics, there is a choice between initiating the dynamics (at $t=0$) with (i) a relaxed wavefunction for the ionized system, and, (ii) an optimized wavefunction for the corresponding neutral system followed by the removal of an electron from its HOMO. 
	In the latter case, the simulation starts with unrelaxed electronic states.
	Fig. \ref{fig:holespbebhlyp2} (top) shows that the localization of the hole in the unrelaxed wavefunction is almost the same for PBE and PBE0, both in the dimer as well as the pentamer chains.
	The results for all chain lengths and for other functionals are presented in supplemental Fig. \ref{fig:holesxcfunctionals}.
	Moreover, as the number of water molecules in the H-bond chain increases, the hole-localization on the first molecule of the chain decreases by the same amount for all functionals. 
	This indicates that the SIE in the unrelaxed wavefunction is non-dominant. 
	It has been hypothesized that such a definition of the initial wavefunction might generate the ``correct dynamics'' even with PBE \cite{doi:10.1021/acs.jpca.8b01259}. Accordingly, rt-TDDFT/PBE benefits from the construction of a ``correct" initial state \cite{doi:10.1021/acs.jpclett.7b01652}.

	On the other hand, when the wavefunction is allowed to relax, the hole-spread increases with increasing chain length (see Fig. \ref{fig:holespbebhlyp2} (bottom), and supplemental Fig. \ref{fig:holesxcfunctionalsrelaxed}).
	The differences between PBE and hybrid functionals are significant.
	In BOMD, such a relaxed wavefunction is always obtained after the convergence of the initial ($t=0$) self-consistency cycle.
	This explains such a large difference between the trajectories for PBE0 and PBE in Fig. \ref{fig:fig13}.
	In PBE, the P(PT) is totally suppressed.
	PBE presents a much larger hole-delocalization than PBE0 at the initial step.
	These results indicate that P(PT) increases with increasing initial photo-hole localization in H-bonded water chains.
	However, our results also show how this localization decreases with increasing chain length -- in fact, this result is general for all XC functionals. 
	It is instructive to analyze how the actual dynamics of the hole in rt-TDDFT, diabatic, dynamics differs between hybrid and non-hybrid functionals. 
	This might help us understand why the proton transfer probability increases despite the decreasing hole localization.
	\begin{figure}
		\includegraphics[width=3.5in]{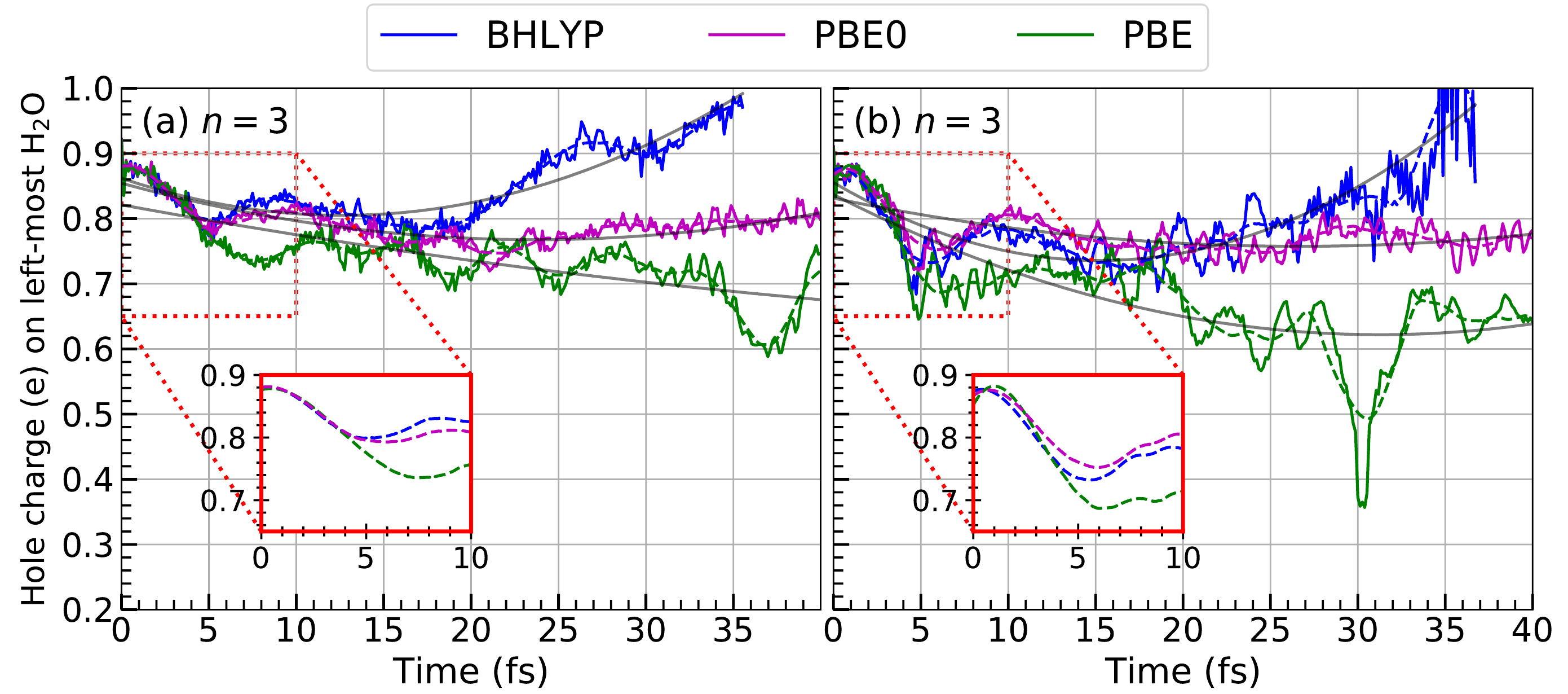}
		\caption{Integrated hole-density localized on the first (left-most) water unit of (H$_2$O)$_n^+$, which loses its electron at $t=0$. Time-evolution of the hole for representative Ehrenfest trajectories: (a) Trimer: successful proton-transfer for all three functionals, (b) Trimer: proton-transfer at PBE0 and BHLYP, but none at PBE. The insets show the short-time behavior of all three functionals in localizing the hole to initiate a first proton-transfer event.}
		\label{fig:trimer_holes}
	\end{figure}
	
	\begin{figure}
		\includegraphics[width=3.5in]{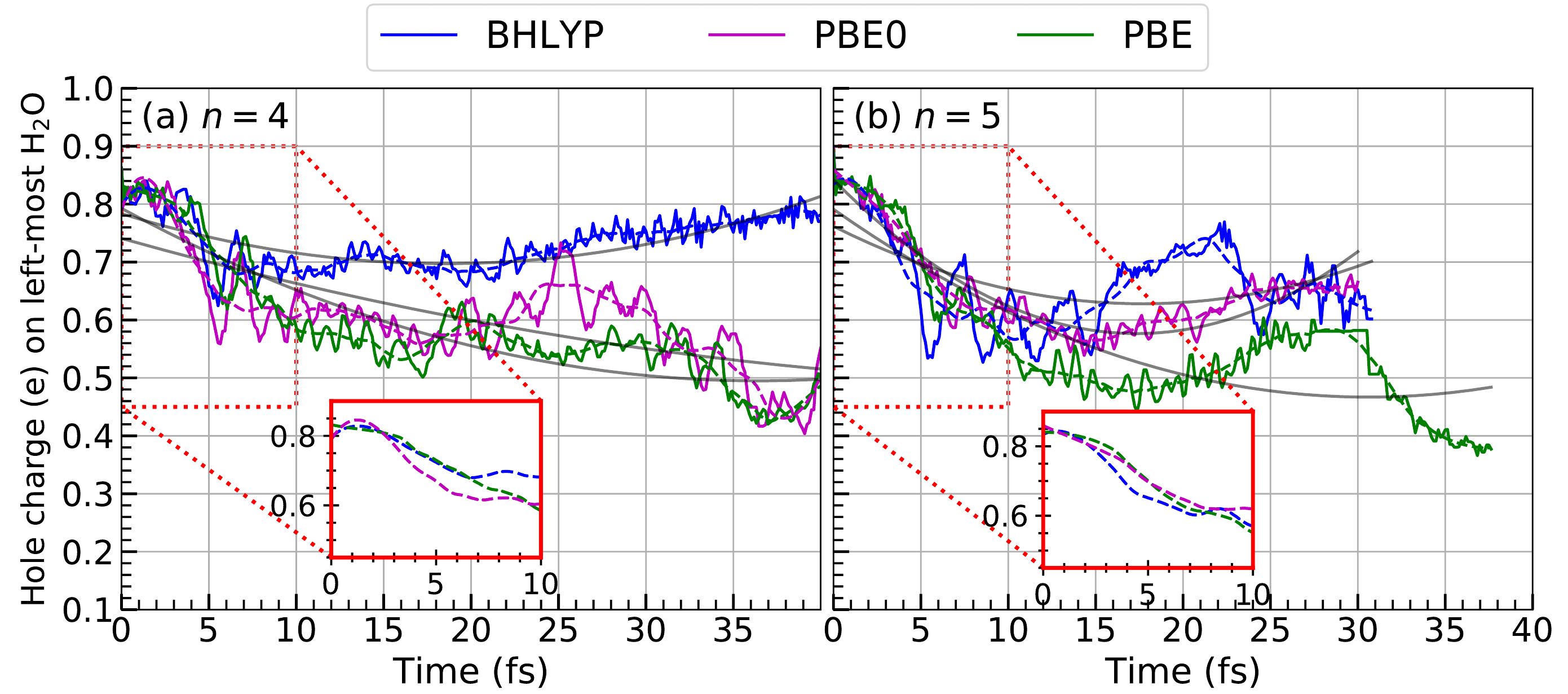}
		\caption{Integrated hole-density localized on the first (left-most) water unit of (H$_2$O)$_n^+$. Time-evolution of the hole for representative Ehrenfest trajectories of ionized: (a) Tetramer, (b) Pentamer.}
		\label{fig:tetrapenta_holes}
	\end{figure}
	
	\subsubsection{Dynamic evolution of the photo-hole in diabatic trajectories}
	We analyze the time-evolution of the hole generated upon removal of an electron from the HOMO of the system at $t=0$. 
	The hole-density at any time instant is estimated by computing the difference between electronic densities of the ground and ionized states for the same geometric configuration.
	We characterize the spatial hole-evolution using:
	\begin{align}
	\Delta\overline{\rho}(x;t) = \iint\!dy\,dz \left[\rho_{\text{ground}}(x,y,z) - \rho_{\text{ionized}}(x,y,z,t)\right] ~,
	\end{align}
	Here, the \textit{x}-axis is chosen to lie along the H-bond chain such that $\Delta\overline{\rho}(x)$ measures the variation of the hole density at H$_2$O positions along the H-bond axis at any given time. 
	The ionized structures and their corresponding TDDFT densities are selected from a single Ehrenfest trajectory.
	Additionally, the ground state densities are computed for these structures using static DFT, which are then subtracted from the respective TDDFT densities.
	We first focus on the hole dynamics in (H$_2$O)$_5^+$ since this system exhibits a high probability of proton transfer.
	Fig. \ref{fig:hole_densities} shows the results for the time-evolution of $\Delta\overline{\rho}(x)$ for a selected Ehrenfest trajectory using GGA/PBE. 
	The relative positions of the evolving molecular species along \textit{x}-axis are also indicated.
	We have confirmed that a majority of the simulated non-adiabatic trajectories describing proton-transfer at the nuclear scale show a similar behavior.
	The hole-evolution begins at $t=0$ when an electron is photo-excited from the HOMO of the system by explicitly changing the occupation number of the state.
	At the start of the simulation, the hole density is predominantly localized on the H$_2$O unit that exclusively donates a H-bond as shown in the isosurface plot of HOMO of the ionized system (inset, Fig. \ref{fig:hole_densities} (a)). 
	A first proton transfer event (from H$_2$O$_1$ to H$_2$O$_2$) completes at $t\sim5$ fs, followed by a secondary proton transfer along the chain (from H$_2$O$_2$ to H$_2$O$_3$) at $t\sim13$ fs, see Fig. \ref{fig:hole_densities} (c), (d).
	While the photohole has significantly delocalized, this does not seem to prevent the occurrence of proton transfers.

	A simple metric for hole delocalization is the hole charge within the photoexcited molecule, obtained by integrating over the water unit at the extreme left of the ionized chain. This is given by
	\begin{align}
	Q_{\text{left}}(t) = \int_{\text{(H$_2$O)$_{\text{left}}$}} dx \ \Delta\overline{\rho}(x;t) ~.
	\end{align}
	The time evolution of $Q_{\text{left}}(t)$, for different XC functionals is shown in Fig. \ref{fig:trimer_holes}, where the integrated hole density is computed at every hundredth step-interval (0.13 fs).
	Fig. \ref{fig:trimer_holes}(a) and (b) show two independent Ehrenfest trajectories for (H$_2$O)$_3^+$, starting from different initial geometric configurations.
	In Fig. \ref{fig:trimer_holes}(a), PBE and PBE0 trajectories complete a first proton-transfer event, while BHLYP provides an additional secondary proton-transfer, however in Fig. \ref{fig:trimer_holes}(b), PBE shows no proton-transfer while both the hybrids PBE0 and BHLYP suggest up to a secondary proton-transfer.
	The insets capture the short-time hole dynamics relevant for the first proton transfer which occurs within 4.61-4.87 fs in the successful cases. 
	In Fig. \ref{fig:trimer_holes}(a), the hole charges at the instant of the first proton-transfer are 0.76 (PBE), 0.78 (PBE0) and 0.79 (BHLYP). Additionally, a secondary proton-transfer occurs for BHLYP at 29.74 fs with a hole-charge of 0.86. 
	In contrast, the values for the trajectory shown in (b) are found to be 0.72 (PBE0) and 0.73 (BHLYP) for the first proton-transfer followed by a secondary proton-transfer occurring around 31.58 fs (PBE0$_{\text{hole}}$=0.77; BHLYP$_{\text{hole}}$=0.82).
	PBE fails to show any proton-transfer and the associated hole-charge decays to 0.66 in the early-time dynamics.
	A point of commonality for all the system sizes $n$, is that initially all the functionals produce nearly identical hole localizations.
	Furthermore, there is not a large and obvious difference between the $Q_{\text{left}}(t)$ of PBE and PBE0.

	We observe that even when there is sufficient delocalization, the electronic charge lags behind the nuclear charge, and its evolution is mediated by the nuclear motion over a prolonged period of time ($\sim$50 fs, which is long when compared to the natural time scale of electronic motion).
	Fig. \ref{fig:tetrapenta_holes} (a) and (b) illustrate the underlying effect of H-bond cooperativity present in longer chains on hole density evolution. The overlapping curves highlighted in the insets show that the cooperative H-bonds in larger $n$ chains reduce the dependence of both hole and proton transfer on the choice of functional.
	After an initial proton transfer reaction, the proton often proceeds to move to the next water molecule in the chain. 
	The nature of this reaction is different from the primary one as both the reactants and the products are inherently different from those found in the primary reaction of interest. 
	In the case of the secondary proton hop reaction, this is initiated
	from an already-formed Zundel complex \cite{Roberts15154}, where the proton hops from the hydronium to the next water unit. 
	At the nuclear level, a secondary proton-hop reaction is improbable in shorter chains. In these structures ($n=2,3$), the hemibonded configuration is still accessible (and favored by PBE) in the simulation -- which suppresses the `proton-transfer' branch and results in a dissociation of the H$_2$O monomers without any proton transfer. 
	On the other hand, we see a significant secondary proton hop dynamics in longer ionized water chains ($n=4,5$), and the probability of such a reaction grows with the number of water molecules in the chain. 
	This explains why, for $n=5$, the probability of a first proton transfer is greater than 90$\%$.
	
	\section{Conclusions}
	We have studied the proton transfer mechanism upon photoionization of H-bonded water molecular chains, as a function of the length of the chain using non-adiabatic rt-TDDFT simulations in the Ehrenfest approximation.
	The goal of this study was to understand how the self interaction error in semilocal (GGA) density functionals influences the proton transfer probability.
	We have shown that for PBE, this probability increases from $13\%$ in the dimer trajectories to $94\%$ in the pentamer trajectories.
	We have also shown that while the probability is largely underestimated in small chains (dimers and trimers) as compared to what hybrid functionals predict, the error is minimal in longer molecular chains. 
	The results indicate that for PBE, the proton transfer is 
	disfavored in cluster sizes of 3 molecules or less, due to the overestimation of the stability of hemibonded geometries
	over proton transfer geometries, which is in turn due to the self interaction error. 
	In longer chains ($n=4,5$ in (H$_2$O)$_n^+$), the increased H-bond cooperativity makes the transition to hemibonded-type geometries less probable. 
	An increasing fraction of the simulated Ehrenfest trajectories exhibit proton-transfer, often along multiple
	water molecules in the chain.

	There is also a clear difference in photohole delocalization in adiabatic, Born-Oppenheimer molecular dynamic simulations, when comparing PBE to hybrid functionals. However, the charge dynamics in rt-TDDFT simulations
	is quite comparable for all the functionals studied in this work.
	This is due to the relaxation of the initial wavefunction of the ionized system.
	The adiabatic scheme (BOMD/PBE) at comparable time scales fails to describe the evolution of the excited system toward a proton-transfer reaction. 
	The inclusion of \textit{non-adiabatic} effects is therefore essential for capturing the proton transfer dynamics.

	The results presented in this paper have important ramifications for photocatalytic water-splitting phenomena occurring on semiconductor surfaces. For example, the presence of H-bonded water chains could improve the photocatalytic activity of the semiconductor. 
	This work shows that in such condensed systems, 
	rt-TDDFT simulations in the Ehrenfest approximation 
	can be carried out using less expensive GGA-type functionals
	without significantly compromising the accuracy of the results.

	\begin{acknowledgments}
		This work was funded by U.S. Department of Energy Award No. DE-FG02-09ER16052. V.S. and M.V.F.S. would like to thank Stony Brook Research Computing and Cyberinfrastructure, and the Institute for Advanced Computational Science at Stony Brook University for access to the high-performance SeaWulf computing system, which was made possible by a \$1.4M National Science Foundation grant (\#1531492).
	\end{acknowledgments}

	\bibliography{refspaper}


	\clearpage
	\clearpage 
	\setcounter{page}{1}
	\renewcommand{\thetable}{S\arabic{table}}  
	\setcounter{table}{0}
	\renewcommand{\thefigure}{S\arabic{figure}}
	\setcounter{figure}{0}
	\renewcommand{\thesection}{S\arabic{section}}
	\setcounter{section}{0}
	\renewcommand{\theequation}{S\arabic{equation}}
	\setcounter{equation}{0}
	\onecolumngrid
	
	\begin{center}
		\textbf{Supplemental Material for\\\vspace{0.5 cm}
			\large Proton-transfer dynamics in ionized water chains using real-time Time Dependent Density Functional Theory\\\vspace{0.3 cm}}
		
		Vidushi Sharma$^{1,2}$ and Marivi Fern\'{a}ndez-Serra$^{1,2}$
		
		\small
		
		$^1$\textit{Department of Physics and Astronomy, Stony Brook University, Stony Brook, New York 11794-3800, United States}
		
		$^2$\textit{Institute for Advanced Computational Science, Stony Brook University, Stony Brook, New York 11794-5250, United States}
		
		(Dated: \today)
	\end{center}
	
	\section{Self-interaction error: more ionized water geometries}
	\begin{figure}[h]
		\includegraphics[width=5.3in]{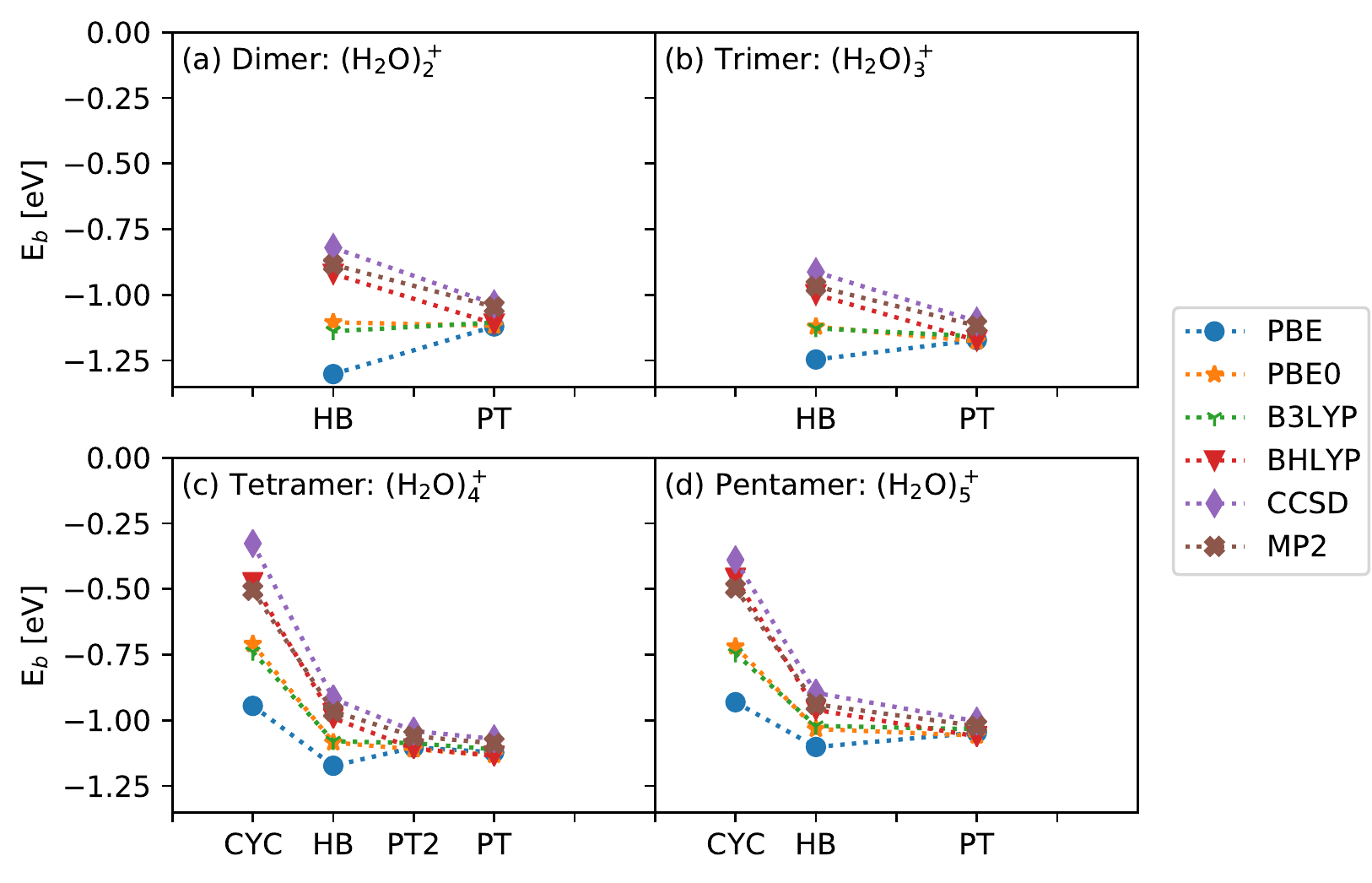}
		\caption{Binding energies of the hemibonded (HB), proton-transferred (PT), cyclic (CYC) and another low-energy proton-transferred (PT2) geometries of the HOMO-ionized water clusters (H$_2$O)$_n^+$; (a) $n=2$, (b) $n=3$, (c) $n=4$ (d) $n=5$. The associated structures for all $n$ are shown in the table below.}
		\label{fig:all_sie}
	\end{figure}
	
	\begin{figure}[h]
		\includegraphics[width=4.1in]{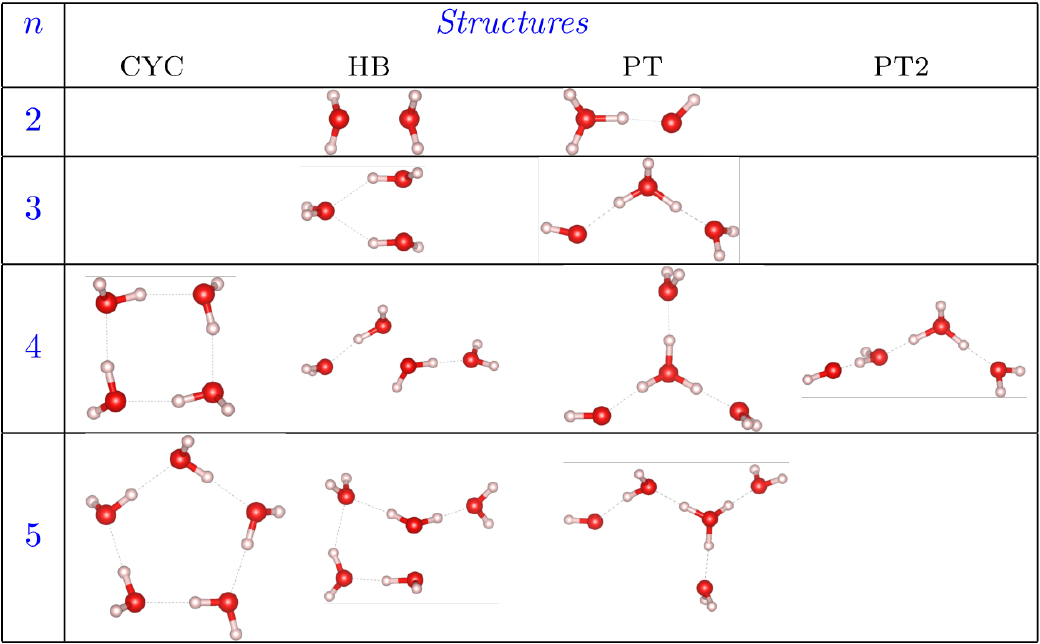}
		\label{fig:sie_structures}
	\end{figure}
	
	\begin{table}
		\caption{\label{tab:bindingenergies}Binding energies of HOMO-ionized (H$_2$O)$_n^+$ structures computed at different levels of theory. All energy values are reported per water monomer for a cluster containing $n$ H$_2$O molecules, $n = (2-5)$. The abbreviations used in the table are HB : Hemibonded, PT : Proton-transferred, CYC : Cyclic, PT2 : (another low energy) Proton-transferred structure.}
		\begin{ruledtabular}
			\begin{tabular}{cccccccc}
				\multicolumn{1}{c}{\textcolor{blue}{$n$}}&\multicolumn{7}{c}{\textcolor{blue}{E$_{b}$ [eV]}}\\
				$ $&$ $&PBE&PBE0&B3LYP&BHLYP&MP2&CCSD\\ \hline
				\multirow{2}{*}{\textcolor{blue}{$2$}}&HB&$-1.302$&$-1.104$&$-1.139$&$-0.920$&$-0.885$&$-0.820$\\
				&PT&$-1.120$&$-1.118$&$-1.105$&$-1.110$&$-1.046$&$-1.035$\\ \hline
				\multirow{2}{*}{\textcolor{blue}{$3$}}&HB&$-1.246$&$-1.121$&$-1.128$&$-0.999$&$-0.967$&$-0.913$\\
				&PT&$-1.172$&$-1.177$&$-1.158$&$-1.173$&$-1.118$&$-1.101$\\ \hline
				\multirow{4}{*}{\textcolor{blue}{$4$}}&CYC&$-0.945$&$-0.710$&$-0.740$&$-0.475$&$-0.506$&$-0.327$\\
				&HB&$-1.173$&$-1.084$&$-1.080$&$-0.991$&$-0.966$&$-0.917$\\
				&PT&$-1.122$&$-1.132$&$-1.111$&$-1.135$&$-1.088$&$-1.071$\\
				&PT2&$-1.103$&$-1.110$&$-1.087$&$-1.108$&$-1.061$&$-1.041$\\ \hline
				\multirow{3}{*}{\textcolor{blue}{$5$}}&CYC&$-0.931$&$-0.719$&$-0.746$&$-0.458$&$-0.497$&$-0.389$\\
				&HB&$-1.101$&$-1.033$&$-1.021$&$-0.960$&$-0.939$&$-0.896$\\
				&PT&$-1.047$&$-1.058$&$-1.034$&$-1.061$&$-1.022$&$-1.004$\\
			\end{tabular}
		\end{ruledtabular}
	\end{table}
	
	\section{Time step testing for HOMO-ionized water dimer: (H$_2$O)$_2^+$}
	\begin{figure}[h]
		\begin{subfigure}{0.45\textwidth}
			\includegraphics[width=3.in]{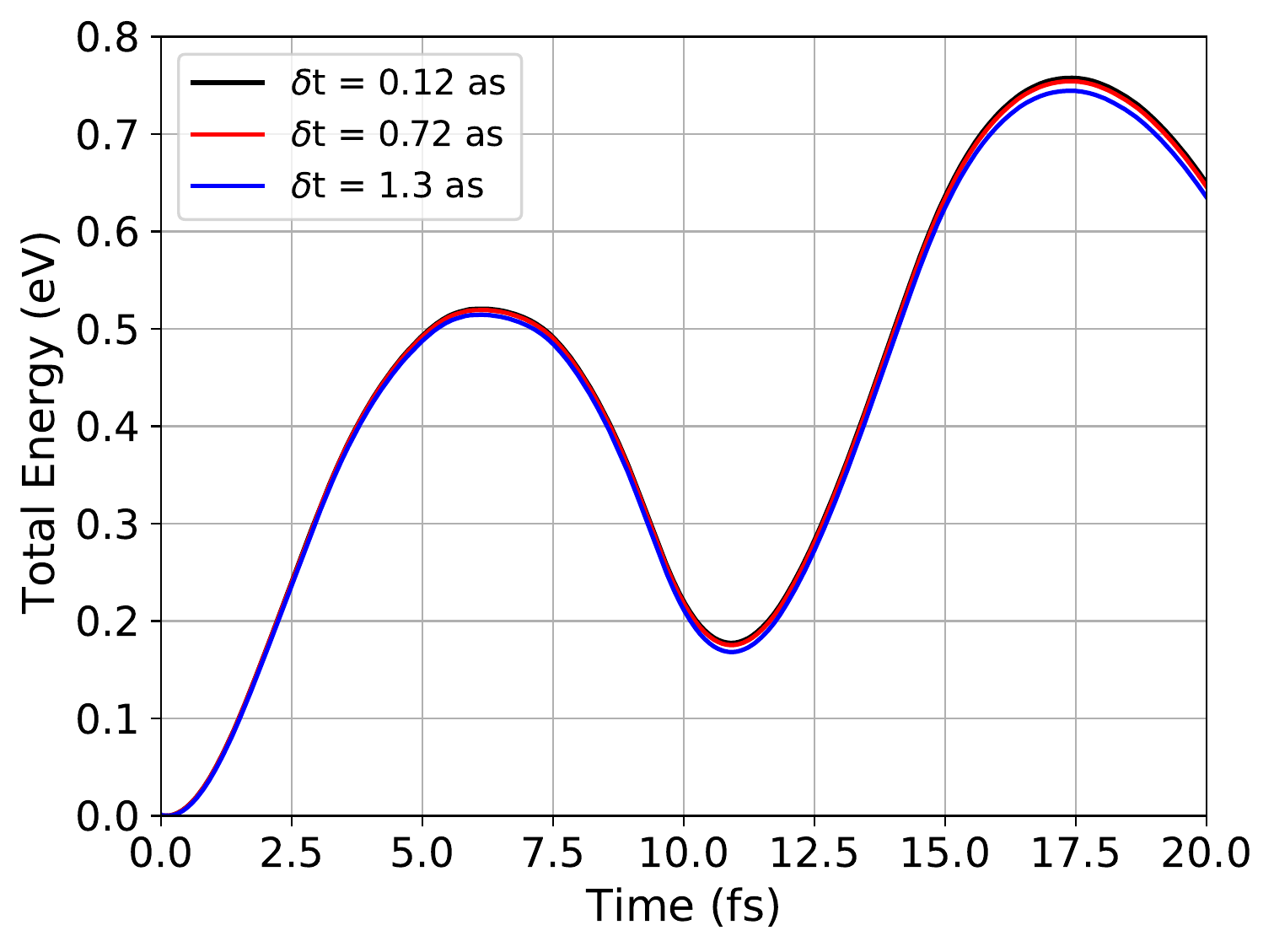}
			\caption{}
			\label{fig:energydimer}
		\end{subfigure}
		\begin{subfigure}{0.45\textwidth}
			\includegraphics[width=3.in]{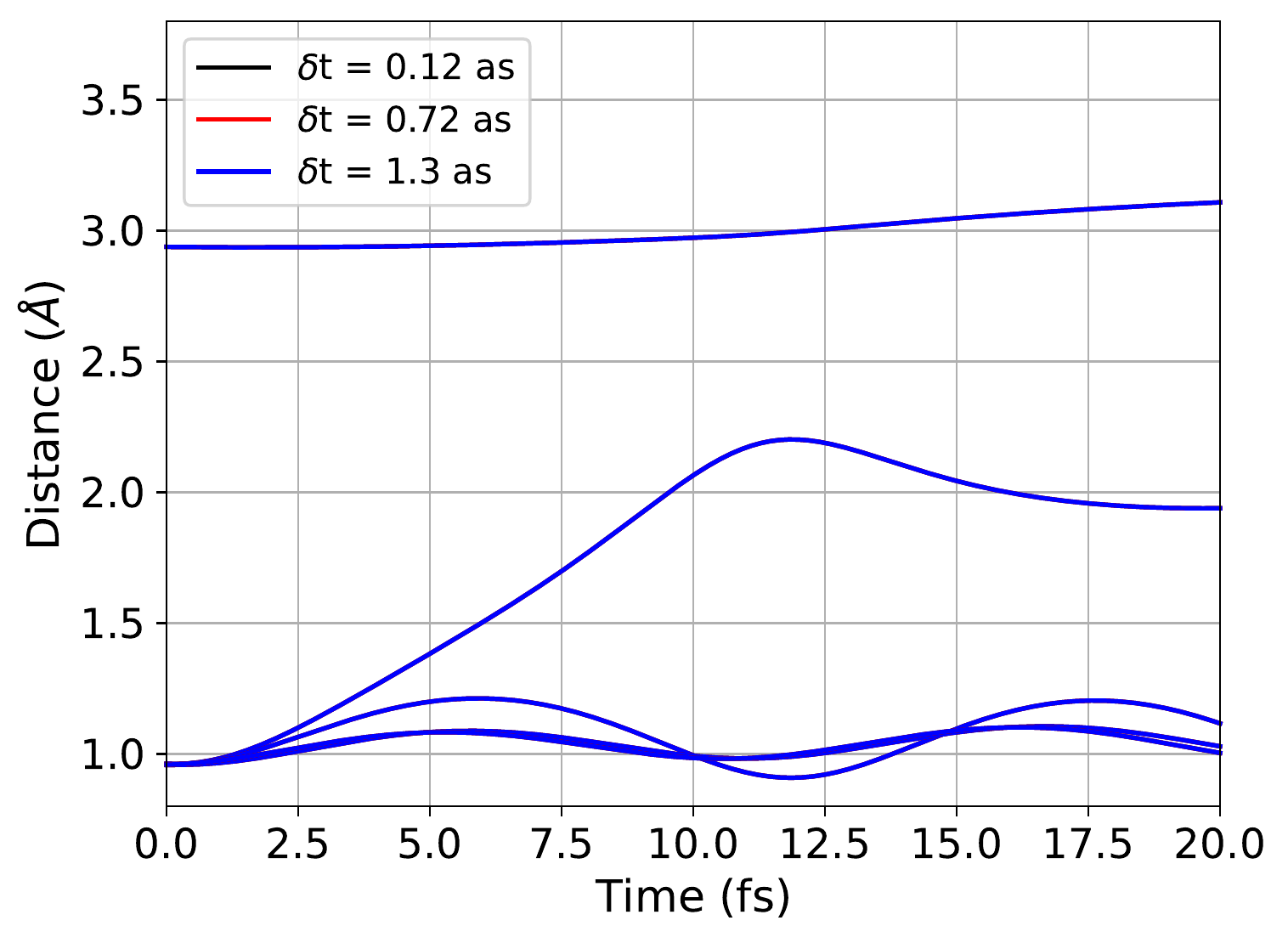}
			\caption{}
			\label{fig:distancedimer}
		\end{subfigure}
		\caption{(a) Energy conservation, and (b) Nuclear dynamics -- all bond lengths and distances with different time-steps in the enforced time-reversal symmetry propagator (ETRS) used for the electronic propagation (in rt-TDDFT/PBE Ehrenfest dynamics).}
	\end{figure}
	
	\newpage
	\section{Trajectory analysis for (H$_2$O)$_n^+$ $n=(2 -5)$}
	\begin{figure}[h]
		\includegraphics[width=5in]{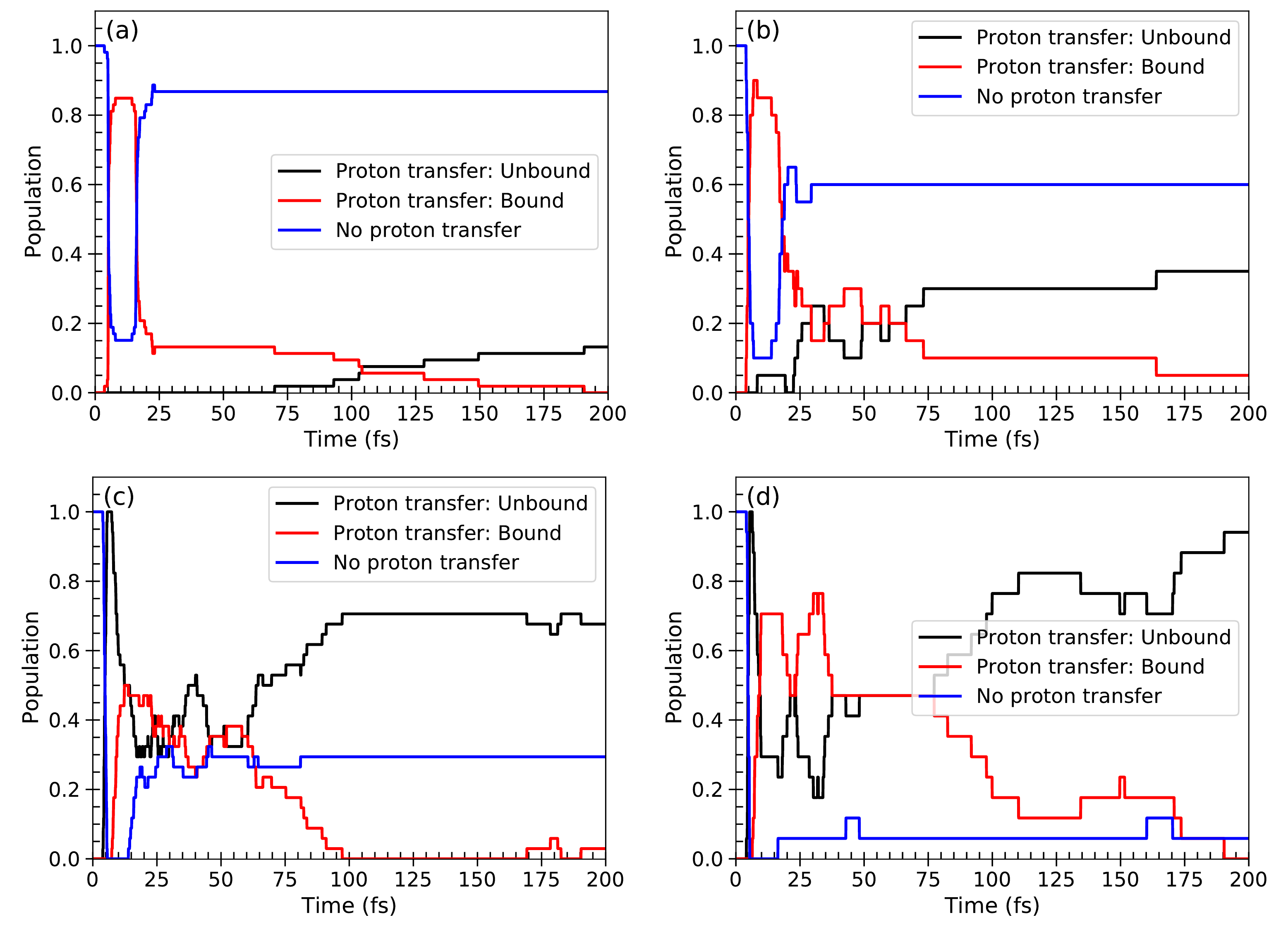}
		\caption{Full population analysis of rt-TDDFT/Ehrenfest dynamics of (H$_2$O)$_n^+$: (a) $n=2$, (b) $n=3$, (c) $n=4$, (d) $n=5$. In the unbound reaction channel, $^{\bm{\cdot}}$OH and H$_{3}$O$^{+}$ dissociate ($d_{\text{O}\cdots \text{O}}> 4.5$ \AA) whereas in the bound channel they continue to interact. Hence, the ``bound" and ``unbound" channels take into account the separation between the $^{\bm{\cdot}}$OH radical formed and the evolving H$_3$O$^+$ ion.}
		\label{fig:SIpopulation}
	\end{figure}
	
	\section{Proton transfer population in ionized water clusters, (H$_2$O)$_n^+$ $n=(2 - 5)$}
	\begin{figure}[h]
		\begin{subfigure}{0.65\textwidth}
			\includegraphics[width=3.5in]{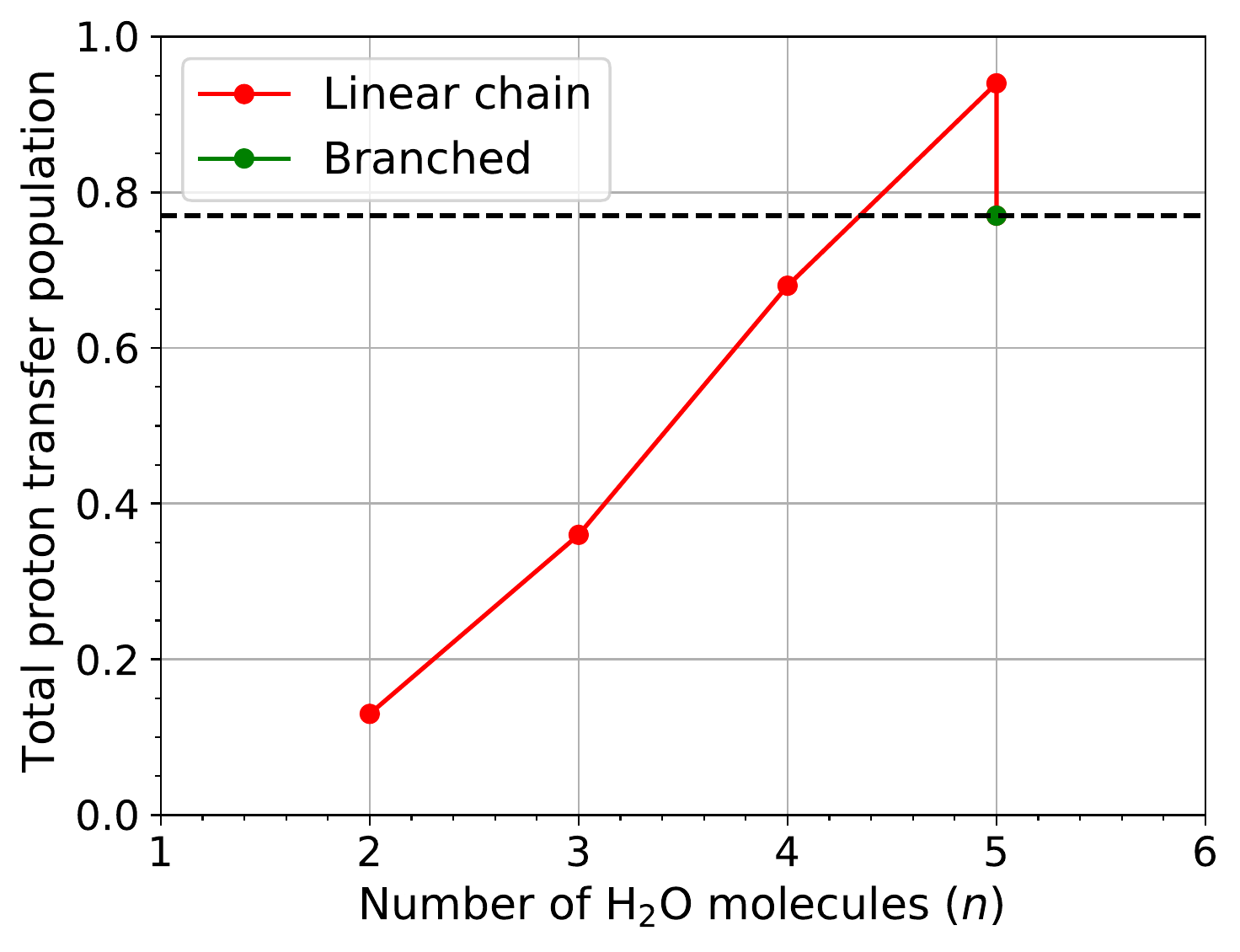}
			\label{fig:total_protontransfer}
		\end{subfigure}
		\begin{subfigure}{0.25\textwidth}
			\includegraphics[width=1.8in]{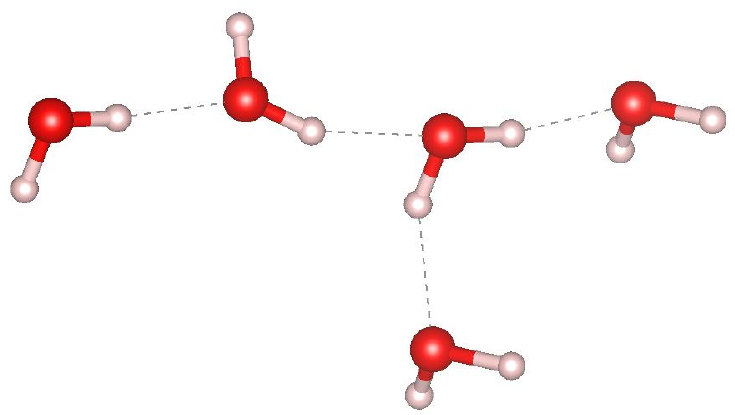}
			\label{fig:pentamer_branch}
		\end{subfigure}
		\caption{Net proton transfer population obtained at the end of simulated rt-TDDFT/PBE Ehrenfest trajectories ($t=200$ fs) for linear (red) and branched (green) ionized water chains, (H$_2$O)$_n^+$. The structure on the right is an example of a ``branched" water geometry consisting of $n=5$ H$_2$O molecules.}
	\end{figure}
	
	\section{Hole density evolution for an ionized trimer chain (H$_2$O)$_3^+$ at PBE and PBE0}
	\begin{figure}[h]
		\includegraphics[width=6in]{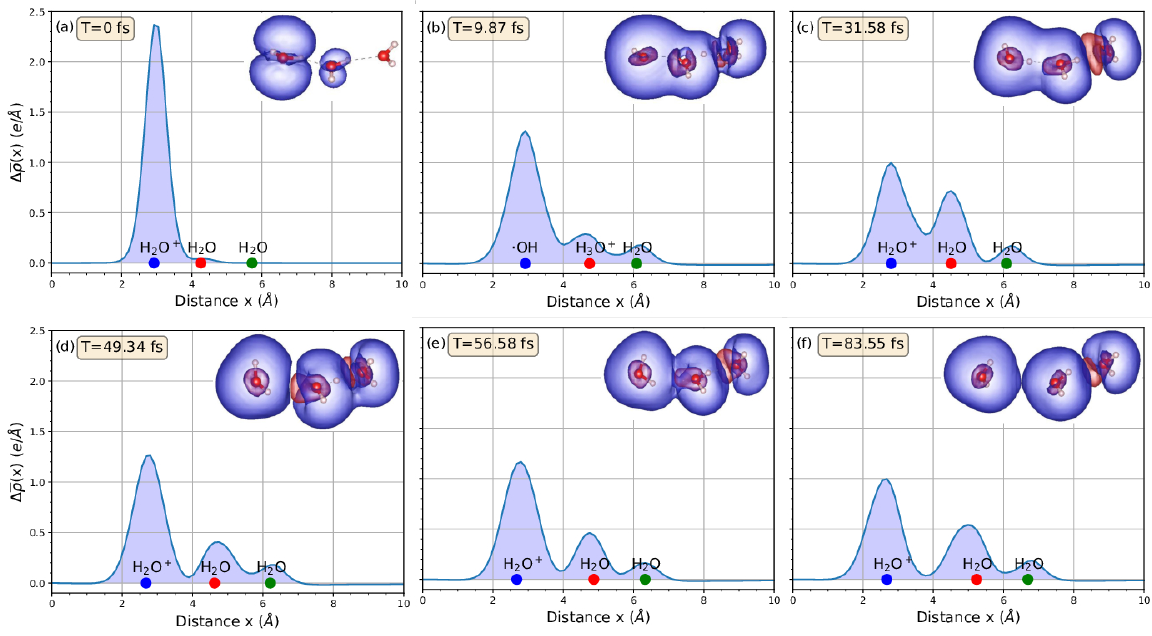}
		\caption{Snapshots of the electronic density differences between ground-state and ionized (+1) configurations obtained at various times showing the time-evolution of hole density - for a single (H$_2$O)$_3^+$ Ehrenfest trajectory at \textbf{PBE}. T= 0 indicates the hole created at the time of ionization of the trimer-water chain. This particular rt-TDDFT/PBE trajectory does not give a proton-transfer as shown in the first row of plots.}
		\label{fig:trimer_pbeholes}
	\end{figure}
	
	\begin{figure}[h]
		\includegraphics[width=6in]{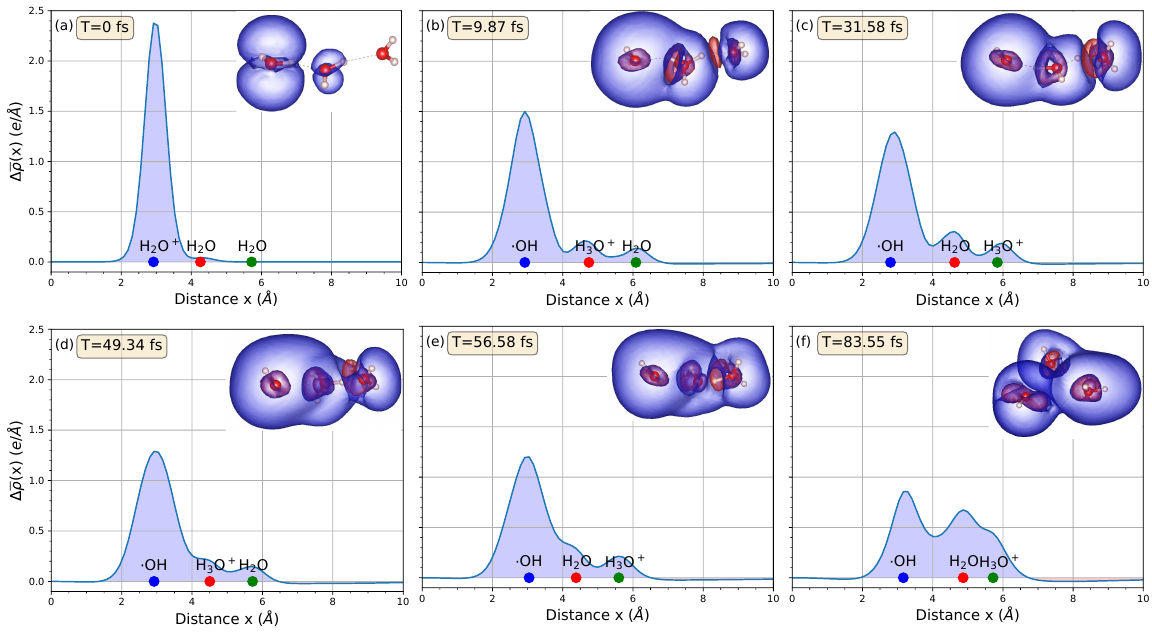}
		\caption{Snapshots of the electronic density differences between ground-state and ionized (+1) configurations obtained at various times to show the time-evolution of hole density - for a single (H$_2$O)$_3^+$ Ehrenfest trajectory at \textbf{PBE0}. T= 0 indicates the hole created at the time of ionization of the trimer-water chain. The rt-TDDFT/PBE0 trajectory starts from the same initial water geometry as in Fig. \ref{fig:trimer_pbeholes} but shows a successful proton-transfer.}
		\label{fig:trimer_pbe0holes}
	\end{figure}

	\begin{table}
		\caption{\label{tab:gradients}Mean gradients of hole charge computed for a simulation time window: $t=[0-10]$, as shown in the insets of Fig. \ref{fig:trimer_holes}, \ref{fig:tetrapenta_holes}.}
		\begin{ruledtabular}
			\begin{tabular}{cccc}
				\multicolumn{1}{c}{\textcolor{blue}{$n$}}&\multicolumn{3}{c}{\textcolor{blue}{Hole charge gradient}}\\
				$ $&PBE&PBE0&BHLYP\\ \hline
				\textcolor{blue}{$3$}&$-0.012\pm0.002$&$-0.007\pm0.001$&$-0.006\pm0.002$\\ \hline
				\textcolor{blue}{$4$}&$-0.025\pm0.002$&$-0.017\pm0.004$&$-0.011\pm0.003$\\\hline
				\textcolor{blue}{$5$}&$-0.029\pm0.002$&$-0.024\pm0.002$&$-0.026\pm 0.003$\\ 
			\end{tabular}
		\end{ruledtabular}
	\end{table}

	\begin{figure}[h]
		\includegraphics[width=7.5in]{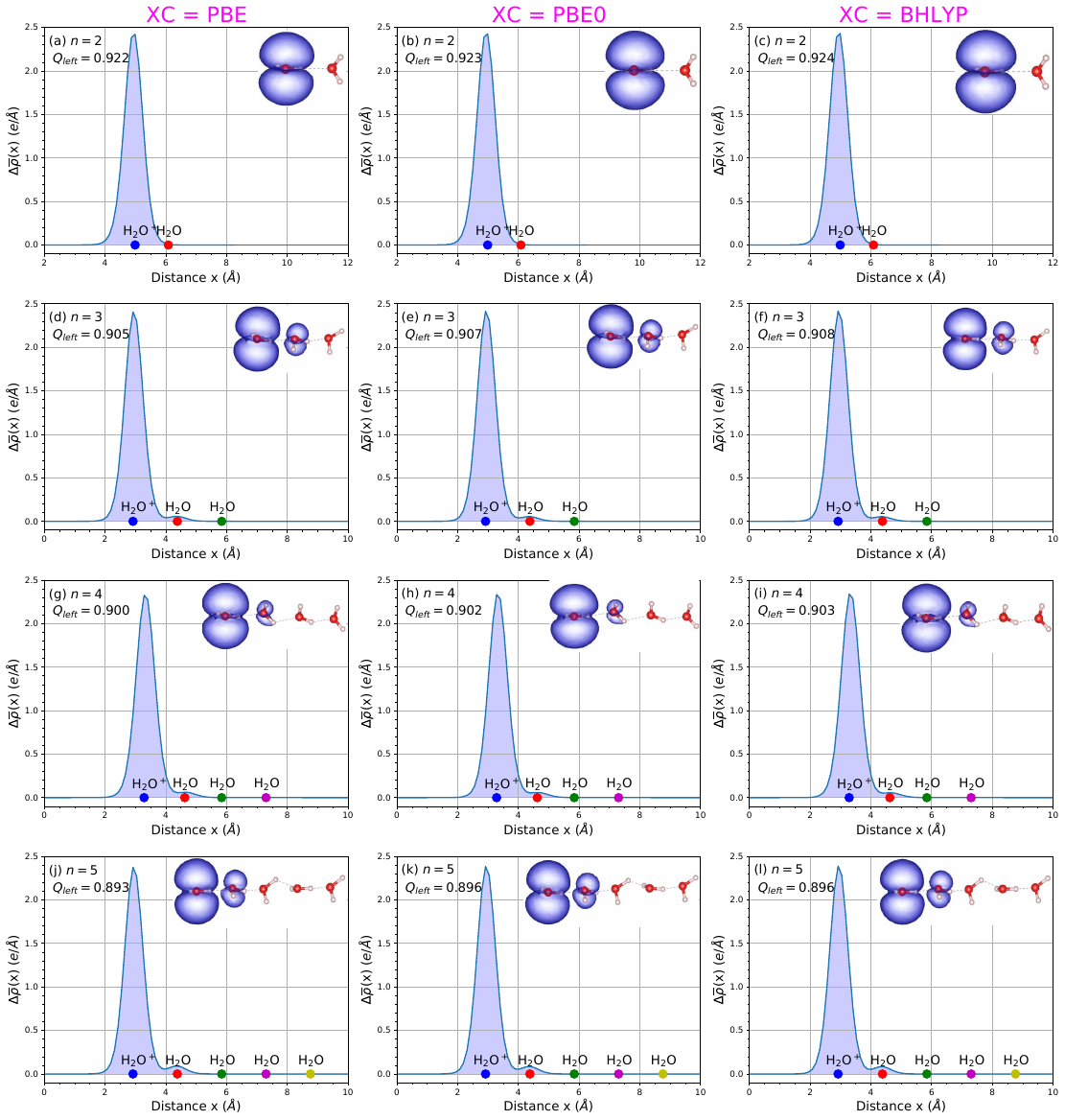}
		\caption{Snapshots of the electronic density differences between ground-state and ionized (+1) configurations depicting the hole densities created at the instant of ionization ($t=0$) of (H$_2$O)$_n$ for $n=(2-5)$, using a given XC functional. The hole densities at $t=0$ are impervious to the choice of the functional.}
		\label{fig:holesxcfunctionals}
	\end{figure}
	
	\begin{figure}[h]
		\includegraphics[width=7.5in]{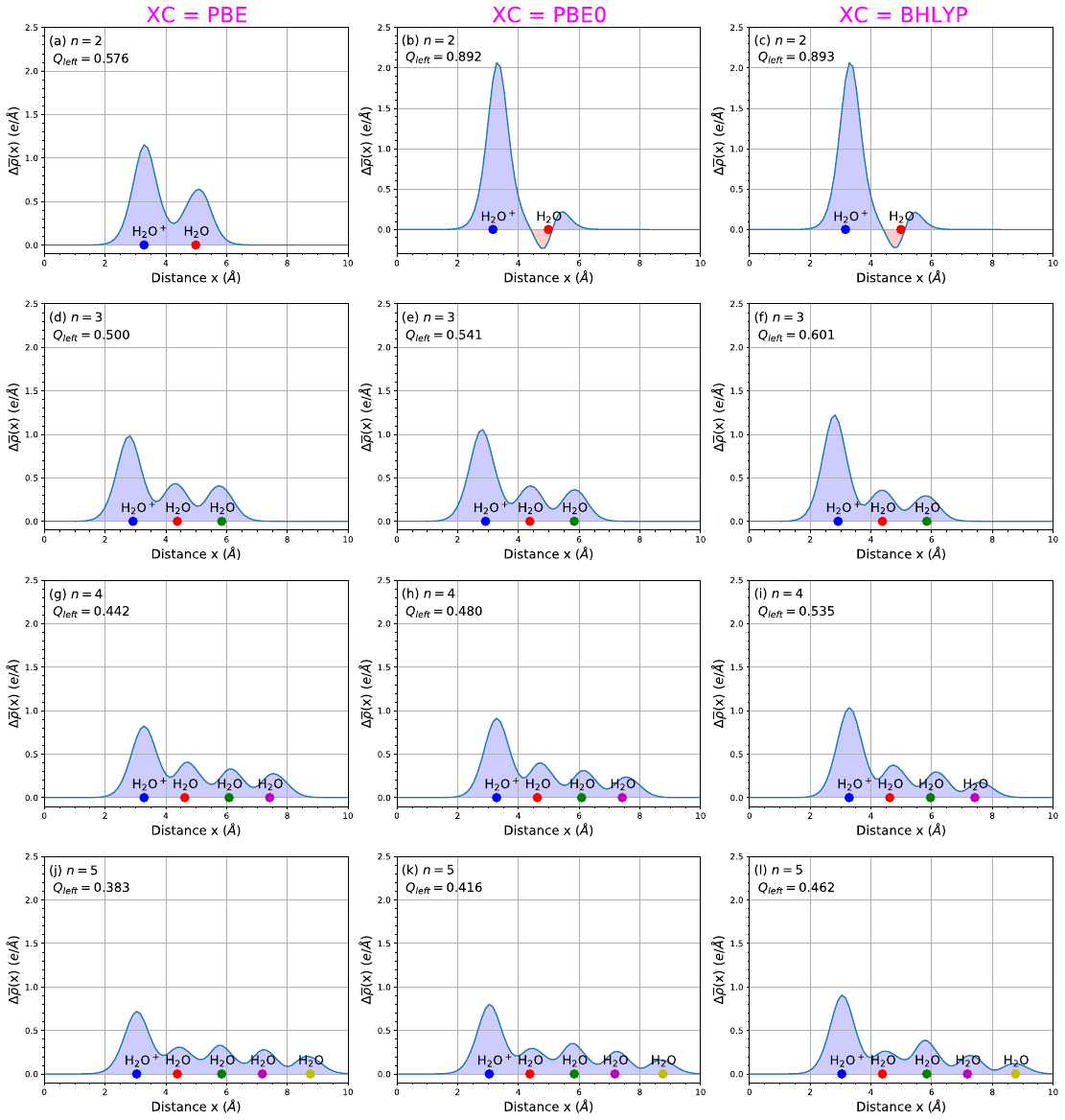}
		\caption{Snapshots of the electronic density differences between ground-state and ionized (+1) configurations at $t=0$, when the ionized wavefunction is allowed to relax using a given (PBE, PBE0, BHLYP) functional. (The ground-state neutral wavefunction is always relaxed throughout this study.)}
		\label{fig:holesxcfunctionalsrelaxed}
	\end{figure}
	
	
\end{document}